\documentclass[a4paper,12pt]{report}
\usepackage{amssymb}
\usepackage{amsmath}
\usepackage{graphicx}
\usepackage{amsfonts}
\usepackage[ruled,linesnumbered,noline]{algorithm2e}
\usepackage{epsfig}
\usepackage{times}
\usepackage{psfrag}
\usepackage{pstricks}
\usepackage{comment}
\usepackage{subfigure}
\usepackage{multirow}
\usepackage{paralist}
\usepackage{mdwlist}
\usepackage{times}
\usepackage[mathscr]{eucal}
\usepackage{lscape}

\usepackage{graphics}
\usepackage{verbatim}
\usepackage{longtable}
\usepackage{rotating}
\usepackage{setspace}
\usepackage{color}

\usepackage{fancyhdr}
\renewcommand{\headrulewidth}{0.5pt}

\fancypagestyle{plain}{%
  \fancyhead{} 
   \renewcommand{\headrulewidth}{0pt} 
}

\begin{document}

\begin{titlepage}
\begin{center}
\vspace*{1in}
{\huge {\textbf{Distributed Data Verification Protocols in Cloud Computing}}}
\par
\vspace{0.3in}
{\large \textbf{Dissertation}}
\par
\vfill
Submitted in partial fulf\mbox{}illment of the requirements
\par
\vspace{0.1in}
  for the degree of
\par
\vspace{0.2in}
{\Large $\mathfrak{ Master\; of\; Technology}$}

\par
\vspace{0.2in}
\textbf{BY}
\par
\vspace{0.2in}
\textbf{\large Priodyuti Pradhan}
\par
\textbf{Reg No.-1106414}
\par
\textbf{Network and Internet Engineering}
\par
Under the guidance of
\par
\vspace{0.2in}
\textbf{\large Prof. R. Subramanian}
\par
\begin{figure}[tbh]
\centering
\includegraphics[scale=0.1]{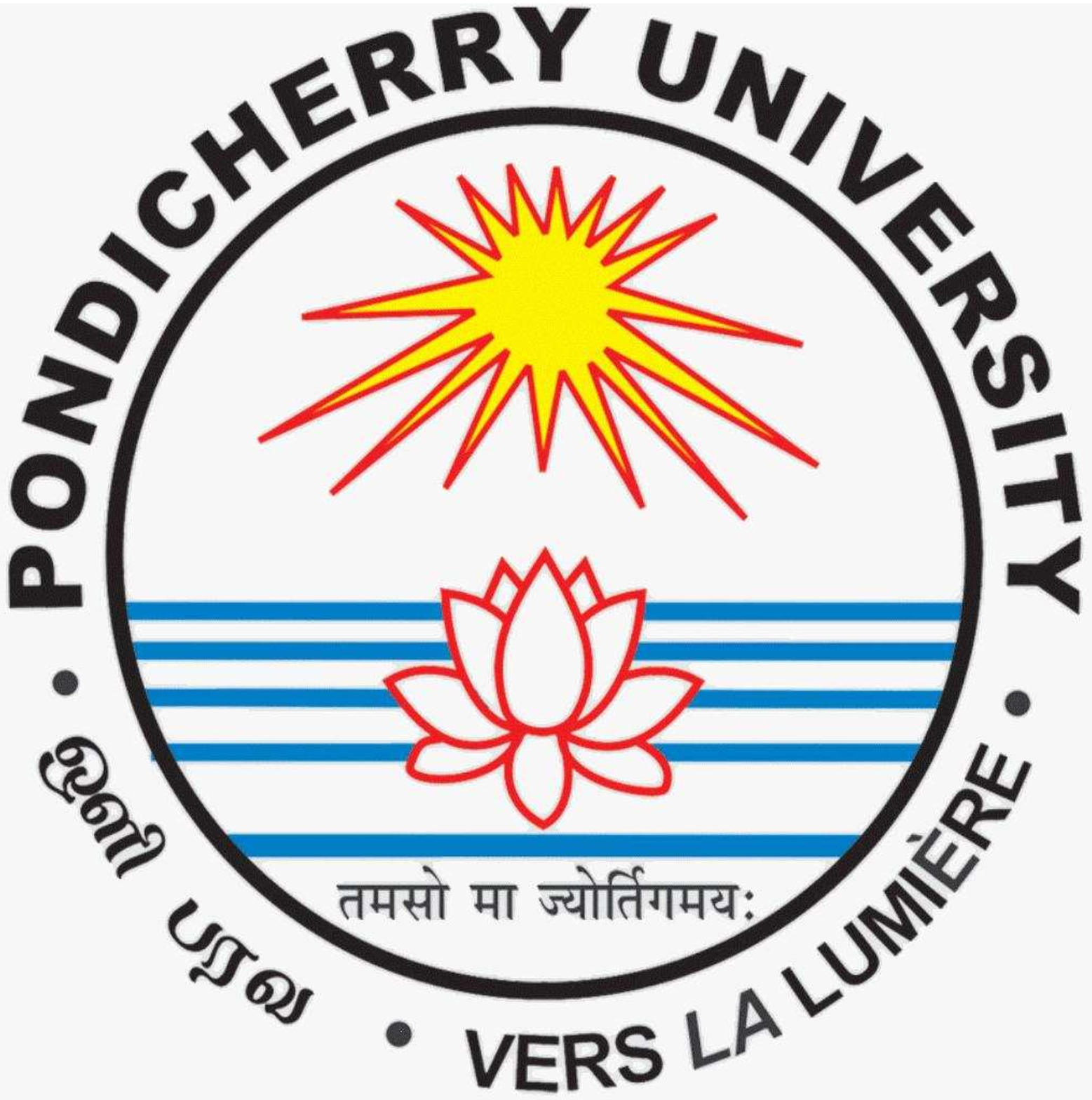}
\end{figure}
\par
$Department\; of\; Computer\; Science\; and\; Engineering$
\par
\vspace{0.4in}
May 2012
\renewcommand{\headrulewidth }{0pt}
\pagenumbering{gobble}
\newpage
\mbox{}
\newpage
\vspace*{1in}
\mbox{}
\vspace{0.2in}
{\Large $\mathcal{ R.S.}$}
\par
{\large $ \mathscr{ DEDICATED\; TO }$}
\par
\vspace{0.15in}
\textit{My Grand Father and My Family}
\newpage
\end{center}
\end{titlepage}


\pagenumbering{roman}
\tableofcontents
\listoffigures
\listoftables

\chapter*{Acknowledgements}

First and foremost, I am very happy to show my immense gratitude and grateful acknowledgement to my beloved guide \textbf{Professor R. Subramanian}, Department of Computer Science, Pondicherry University, for his excellent guidance, constant encouragement and centering efforts, which helped me to complete my directed study report. My gratitude also extends to \textbf{P. Syam Kumar}, Research Scholar, Department of Computer Science, Pondicherry University, for his invaluable suggestion, mentorship, and assistance during the project. 

I express my sincere thanks to \textbf{Dr. P. Dhavachelvan}, Professor and Head, Department of Computer Science, Pondicherry University, and \textbf{Professor Gautam Mahapatra}, Asutosh College, University of Calcutta, for his kind advices and constant encouragement during the course of this project work.

It is my immense pleasure to say thanks to all the \textbf{faculty members}, \textbf{classmates}, and \textbf{non-teaching staff} of the Department of Computer Science, for all the assistance they rendered so willingly, to help me in completing my project report.

Finally, I am extremely thankful to my \textbf{parents} and \textbf{friends} for giving me the moral support in doing all things.

\par
\vspace{1in}
\hspace*{3.6in}\textbf{Priodyuti Pradhan}
\begin{abstract}
Recently, storage of huge volume of data into Cloud has become an effective trend in modern day Computing due to its dynamic nature. After storing, users deletes their original copy of the data files. Therefore users, cannot directly control over that data. This lack of control introduces security issues in Cloud data storage, one of the most important security issue is integrity of the remotely stored data. Here, we propose a \emph{Distributed Algorithmic} approach to address this problem with publicly probabilistic verifiable scheme. Due to heavy workload at the Third Party Auditor side, we distributes the verification task among various SUBTPAs. We uses \emph{Sobol Random Sequences} to generates the random block numbers that maintains the uniformity property. In addition, our method provides uniformity for each subtasks also. To makes each subtask uniform, we uses some analytical approach. For this uniformity, our protocols verify the integrity of the data very efficiently and quickly. Also, we provides special care about critical data by using \emph{Overlap Task Distribution Keys}.
\end{abstract}

\pagenumbering{arabic}

\chapter{Introduction}
\label{ch:intro}
Cloud computing refers to the latest computing technology that enables utility based computing~\cite{birdseye}, i.e. pay by use rather than the ownership of computing resources. The utility part can be hardware, system software or application software that can be accessed from anywhere and used anytime. Typically the interface used for accessing the utility is web based. Cloud computing is a result of evolution and convergence of several independent computing trends like utility computing, virtualization, distributed and grid computing, elasticity, Web2.0, service oriented architectures, content outsourcing and internet delivery. Thus, the cloud can be viewed as an extension of the Internet, wherein opportunities for using large-scale distributed computing infrastructure are being explored for tangible solutions to applications relevant to society and its businesses. Cloud computing, as defined by the National Institute of Standards and Technology (NIST), covers the most comprehensive vision of the cloud computing model: ``Cloud computing is a model for enabling convenient, on-demand network access to a shared pool of configurable computing resources (for example, networks, servers, storage, applications, and services) that can be rapidly provisioned and released with minimal management effort or service provider interaction''~\cite{pallis}. Thus, cloud computing is a computing paradigm that abstracts many of the computational, data and software functionalities needed by a community into a virtual, remote and distributed environment. The term cloud refers to both the resources and the associated services that provide effective utilization and remote access of the resources.

\section{Cloud Computing}
One of the core concepts in cloud computing that makes it an attractive paradigm is virtualization. By virtualization of the entire hardware, software, and network stack, cloud services provide a virtual environment of almost limitless capabilities to the user providing the flexibility to use resources of much larger magnitude than what is actually available. The cloud model promotes availability and is composed of five essential characteristics:
\begin{enumerate}
\item[a.]\textbf{On-demand self-service:} A cloud user can locate and launch a cloud service without any third party help.
\item[b.]\textbf{Broad network access:} Ubiquity of service access from any access device like laptop, mobiles, etc., and from anywhere.
\item[c.]\textbf{Resource pooling:} Same resource can potentially be used by simultaneous as well as many different users.
\item[d.]\textbf{Rapid Dynamicity:} As the demand for the service increases, so does the availability of resources to support the demand. Similarly, as service demand decreases, unused resources are released.
\item[e.]\textbf{Measured service:} A service is charged by its usage and hence measured for its usage as against the current models where ownership cost is associated with its use. 
\end{enumerate}

A cloud can be designed to deliver three service models, namely,
\begin{enumerate}
\item[1.]\textbf{Infrastructure as a Service (IaaS) cloud:} This model provides the consumer with the capability to provision processing, \emph{storage}, networks, and other fundamental computing resources, and allow the consumer to deploy and run arbitrary software, which can include operating systems and applications. The user, in this model, can demand, acquire and use resources in the form of CPU cycles or \emph{storage space} dynamically. Amazon Web Services is an example of infrastructure as a cloud service. In this model the cloud user gets the hardware resources as a service, over which he needs to deploy the system and application software meeting his use.
\item[2.] \textbf{Platform as a Service (PaaS) cloud:} This model provides the consumer with the capability to deploy onto the cloud infrastructure, consumer created or acquired applications, produced using programming languages and tools supported by the provider. The consumer does not manage or control the underlying cloud infrastructure including network, servers, operating systems, or storage, but has control over the deployed applications. Google App Engine and MS-Windows Azure are examples of cloud platform as a service.
\item[3.]\textbf{Software as a Service (SaaS) cloud:} A complete user application, offered as a service, forms the cloud software as a service. Google Docs, SalesForce, Zoho are some examples of this cloud service model.
 
\end{enumerate}
Further, clouds can be deployed as~\cite{addcloud}:
\begin{enumerate}
\item \textbf{Private cloud:} The Cloud infrastructure is operated for a private organization. It may be managed by organization or a third party, and access is restricted to the owner or organization.
\item \textbf{Community cloud:}The Cloud infrastructure is shared by several organizations and supports a specific community that has communal concerns. It may be managed by the organizations or a third party, and access by the members forming a community based on common interest and use.
\item \textbf{Public cloud:} The Cloud infrastructure is made available to the general public or a large industry group and is owned by an organization selling Cloud services.
\item \textbf{Hybrid cloud:} The Cloud infrastructure is a composition of two or more clouds (private, community, or public) that remain unique entities, but are bound together by standarized or propritary technology, that enables data and application portability.
\end{enumerate}

Cloud computing also places high emphasis on seamless access through easy-to-use interfaces and on-demand provisioning of resources, aspects that are important for easy adoption of clouds, and effective resource and cost management. Typical cloud middleware components also provide services related to resource discovery, management, mapping, monitoring, replication, accounting, virtualization, problem solving environments, reliability and \emph{security}. While a cloud is yet another large scale distributed systems setup, it is quite different from the traditional distributed systems from the perspective of resource access, ownership and usage. Clouds promote the use of self-service with an on-demand usage model. Thus, the user has the freedom to choose required services and only pay for its usage. This is different from current practices wherein large data-centres need to be owned, for using. The pay-by-use pattern has scope for significant reduction in the total cost of ownership for any organization that is intending to use the cloud. At the same time, clouds promote better commercial opportunities for the providers by allowing optimized usage of resources due to \emph{sharing by different users}.

\subsection{Usage of Cloud Computing}
The key motivators for the cloud computing model~\cite{birdseye} are its features like \emph{availability} (anywhere and anytime), \emph{elasticity} (increase or decrease service capacity), \emph{pay-as-you-go} (utility), and \emph{reduction in cost of ownership} for the compute resources. Cloud computing is highly useful in many scenarios in scientific, administrative (governance), and commercial applications. Cloud computing infrastructure at the national level can address problems of diverse nature. These problems can be related to e-governance applications including archiving documents, sharing information about national policies, rules and rights, propagating education material, managing health records, processing agricultural information, land documents, urban planning, traffic control and coordination etc. Scientific
applications including nanoscience, bioinformatics, climate and weather modeling, molecular
simulations, earthquake modelling, homeland security, surveillance, reconnaissance, remote
sensing, signal and image processing can also be addressed effectively using cloud computing. The \emph{storage or data cloud} will act as a repository of data belonging to different domains and service data requests from the users and computational resources in the computational cloud. E-governance applications like maintaining health records, UID information, bank and property documents, and voting records of about one billion people can lead to huge voluminous data of many exabytes. Utility applications like maintaining digital libraries of books and journals, and archives related to different information can lead to data explosion. Further, close knit communities that can share vital information of mutual interest through clouds can be formed. Hence, data storage services from the Clouds is a very essential for storing large volume of data. Because, without buying any storage devices users can dynamically and distributically store their data into remote Cloud Servers for usage and without storing data in remote places users can not achieves the Cloud Computing facility.

\section{Various Issues in Cloud Computing}
Besides all these diverse usage of Cloud Computing, it needs to address many issues, because this computing is based on the \emph{Distributed Control and Distributed Data} paradigm. Here, we mention some well known issues, such as flexible architectural solutions, efficient resource virtualization techniques (CPU, storage, and link virtualization), performance modeling and optimization, modeling and development techniques for Cloud-based systems and applications, reliability modeling, techniques and policies for ensuring security and privacy, and many others. In addition, various related issues in the development areas of smart devices (phones, tablets, and the like) include mobile and roaming access, mobile and dynamic applications, incorporation of modern wireless and cellular technologies into cloud paradigm, and the development of new virtualization, scheduling and
transport schemes to achieve energy saving and enable green computing to become an intrinsic part of the Cloud. 
The survey conducted by the International Data Corporation (IDC)~\cite{idc} on cloud computing services during august 2008/2009 has revealed that security is the biggest concern.

\section{Security Issue}
Security  is  the  serious concern,  because  most of the clients  of Cloud Computing  worries about their business information and critical IT resources in the Cloud Computing system which are  vulnerable  to  be  attack~\cite{takabi}. Although Cloud Computing does not have any new technologies, its characteristics, service models and deployment models raise new security issues such as \emph{Data Storage Security}, Infrastructure Security, Virtualization Security, Software Security, Platform Security, Privacy etc.  Security implementations will require additional monetary resources to implement.   Service Level  Agreements  (SLAs)  with  cloud providers  are  less  robust  than the expected requirements  for  a  company  providing  IT  services.  Governance and security standards in this regard are currently lacking.  Thus, we need to have efficient and effective methods require handling these security issues.

\subsection{Data Storage Security}
Data storage security is an important security issue in the cloud computing because data is the main things, is also significant aspect of Quality of Service (QoS). Cloud data storage belongs to IaaS which allows clients to move their data from local computing systems to the Cloud. By moving their data into cloud, the clients can be relived from the burden of \emph{management and maintenance} of data locally. Amazon’s Elastic Compute Cloud (EC2) and icloud are well known examples of cloud data storage. This new data storage service also brings about many challenging design issues, which have a profound inf\mbox{}luence on the security and performance of the overall system. The same information security concerns are associated with this data stored in cloud: Confidentiality, Integrity, and Availability. These security issues arise due to the following reasons:

\begin{enumerate}
\item The data loss incidents could happen in any infrastructure no matter what degree of reliable measures the cloud service provides would take.
\item Sometimes, the Cloud Service Providers (CSPs) may be dishonest and they may discard the data which has not been accessed or rarely accessed to save the storage space. Moreover, the CSP data may choose to hide data loss incidents (due to management failure, hardware failure, corrupted by outside or inside attacks etc.).
\item Clouds use the concept of ``multi-tenancy'' where by multiple clients data is processed on the same physical hardware. Hence, unauthorized users may deletes some part of the files resulting loss of integrity. 
\end{enumerate}

Some recent data loss incidents are the sidekick cloud disaster in 2009 and the breakdown of Amazon’s Elastic Computing Cloud (EC2) in 2010 are in~\cite{wangtowards, wangpublic} respectively.
The clients need to have strong evidence that the cloud servers still possess the data and it is not being tampered with or partially deleted over the time and cannot accessed by the unauthorized users. Because, the internal operation details of the CSP may not be known to the clients. 
Encrypting the data before storing in cloud can handle the confidentiality issue. However, verifying integrity of outsourced data is a difficult task without having a local copy of data or retrieving it from the server. Due to this reason, the straightforward cryptographic primitives such as Hashing, Signature schemes for data Integrity and Availability are not directly applicable. Also, it is impractical for the clients to download all the stored data in order to check its integrity, because this would require an expensive I/O cost and communication overhead across the network. Therefore, it is desirable to check the integrity of the remotely stored data files without downloading all the files and in a periodic manner. Hence, it reduces the communication, computational, and storage overhead.

\section{Goal of This Project}
Storing data files into the Cloud storage gives better benefits to the large Enterprises as well as individual users because they can dynamically increases their storage space without buying any storage devices. In addition, users can access the remotely store data files anytime and from anywhere, and also gives permission to shares these data to the authorized users. Besides all of these advantages of data outsourcing, there are some security risks due to store data files into remote servers and one of the most important is checking the integrity of the remotely stored data.

The integrity of the data may be lost due to internal failure (disk failure) on the server side \cite{diskfailures}, or external attacks due to unauthorized users may deletes or edit some part of the files \cite{wangtowards} because Clouds uses the concept of \emph{multi-tenancy} where multiple users processes run on the same physical hardware~\cite{outlook}. Even sometimes, to increase the profit, the CSP may deletes some rarely accessed files~\cite{pdp}. Thus, by keeping huge amount of data into the Clouds, user cannot ensures the integrity of the stored data~\cite{selfdata}. Therefore, user needs some scheme to check the integrity of the outsourced data without downloading the whole files and in a periodic manner.

Recently, many researchers have focused on the problem of remote data security and proposed different protocols~\cite{pdp,wangtowards,wangpublic,audit,shyamecc,shyamrsa,p16} to verify the integrity of remote data stored in distributed storage systems without having a local copy of data based on Remote Data Checking (RDC) protocol. In RDC protocol, the client generates some metadata.  Later, the verifier stores the pre-computed values and sends challenge to the server through Challenge-Response Protocol. Upon receiving a request from the verifier, the server computes a response and returns to the verifier. Then, the verifier checks whether the data has been modified or not. To reduce the computational cost and some security purposes clients uses help from the Third Party Auditor (TPA). Where, TPA act as a verifier and do all the verification related works.
  
In the prior work~\cite{pdp,wangtowards,wangpublic,audit}, uses Pseudo Random sequence to verify the file blocks, but due its nonuniform nature, the error block detection probability is less and also takes more time than Sobol Random sequence. Recently, in~\cite{shyamecc,shyamrsa} uses Sobol Random sequence, to verify the integrity of the data, and ensures strong integrity. But all these above work uses single Third Party Auditor(TPA) for verifying the integrity of the data. Here, we uses multiple TPAs termed as SUBTPAs to check the integrity of the outsourced data.

In case of single Auditor verification protocol, One-Third Party Auditor(TPA) receives the request from the Client and performs the verification task. Thus, in this single Auditor system if TPA system will crash due to heavy workload then all the verification process will be abort. On the other hand, during verification process the network traffic will be very high near the TPA organization and may create network congestion. Thus, the performance will be degrading in this scheme. Here, we device a distributed verification scheme, where the Main TPA will distributes the verification task \emph{uniformly} among many number of SUBTPAs. We have shown our proposed model in Figure~\ref{fig:audit}. 
\begin{figure}[!htbh]
\centering
\includegraphics[scale=0.6]{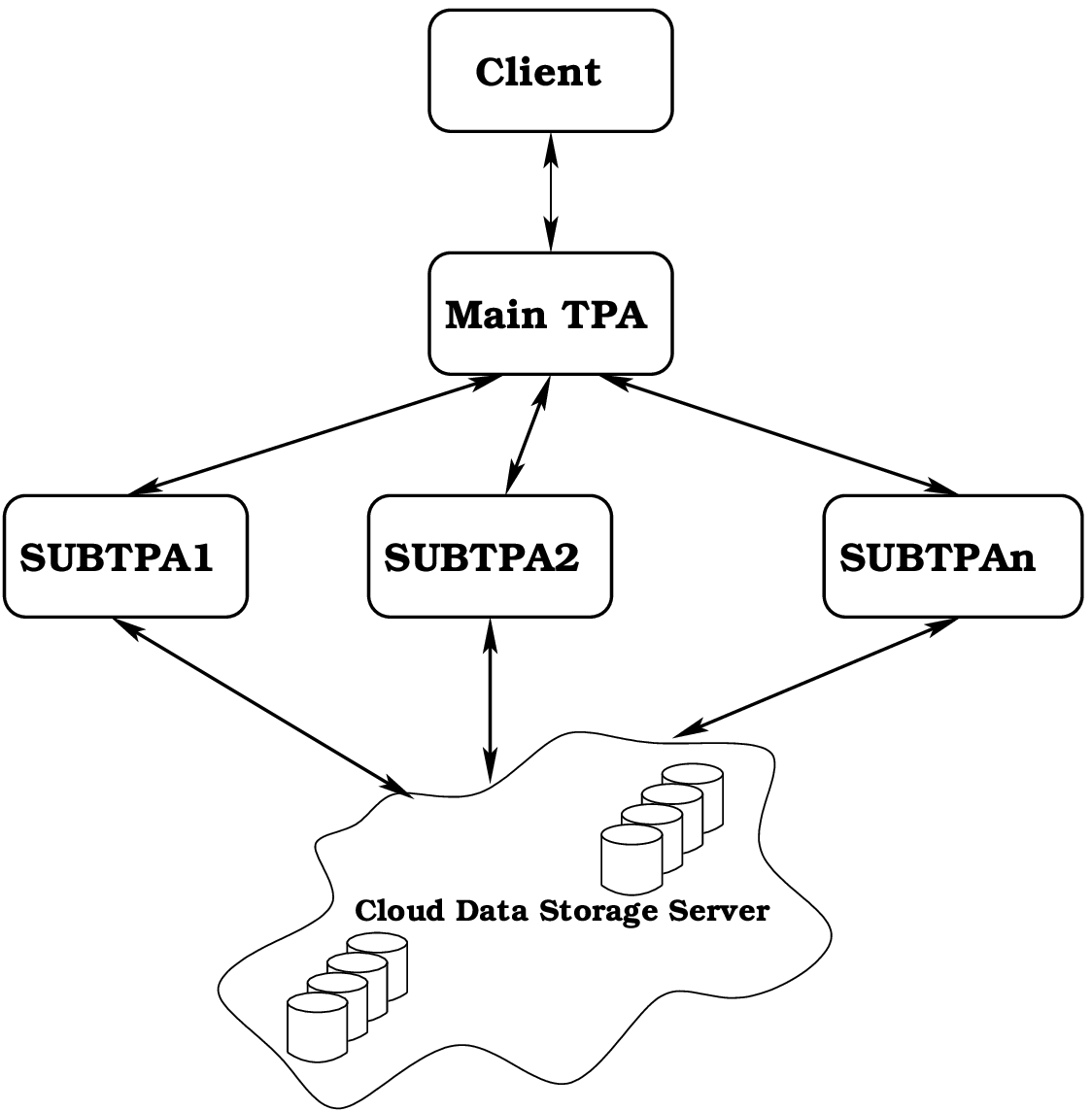}
\caption{Block Diagram of Distributed Audit System Architecture}
\label{fig:audit}
\end{figure}
Our aim is to achieve a general distributed verification protocols where any existing RSA or ECC based protocol will work but in a distributed manner.

\section{Contribution of This Project}
In our protocol, the Client/Main TPA act as a Coordinator and all the SUBTPAs are workers under the Coordinator. Hence, all the SUBTPAs will perform this verification tasks concurrently and gives the verification result to the Coordinator. Thus, concurrency increases the performance of this scheme. In addition, we uses \emph{Sobol Random number generator} [4] instead of general Random number generator to generates the random file block numbers being verify. After that distributes the generated sequence among SUBTPA in a uniformly manner.

For distribution generated Sobol Random Sequence we uses two different approach, one is simple partition based method that satisfy the \emph{Equivalence Relation} property and in second approach we uses random mask called as Task Distribution Key ($\mathcal{TDK}$) for each SUBTPA to interpret the subsequence from the sequence. In addition, we provides verification unifrormity for each SUBTPA by adjusting the $\mathcal{TDK}$ length when necessary. We decides the $\mathcal{TDK}$ length based on the number of SUBTPAs. Here, we uses two types of $\mathcal{TDK}$ one is Non-Overlapping and another is Overlapping $\mathcal{TDK}$. We uses Non-Overlapping $\mathcal{TDK}$ for general applications and for checking the critical data uses Overlapping $\mathcal{TDK}$. In addition, we consider about network bandwidth and workload at the Cloud Serverside and reduce by sending 10\% challenge at a time.

We have shown that, our distributed verification protocols detect errors very fast and also the number of errors detected by each SUBTPA is uniform that means each SUBTPA will detect approximately same number of errors. This is happen due to use Sobol Random Sequence. In addition our protocol detect number of errors depens on the size of generated sequence that means if generated sequence length is 20\% of the total file blocks and there are 1\% error blocks then detects 20\% of the 1\% error blocks and this ratio is true for all cases. Therefore, we can summarizes as follows::
\begin{enumerate}
\item We device distributed Verification protocols based on simple partition based and random Task Distribution Key($\mathcal{TDK}$) based approach.
\item We uses two types of $\mathcal{TDK}$, \emph{Non-Overlapping and Overlapping} based on the applications.
\item Here we decides the $\mathcal{TDK}$ length based on the number of SUBTPAs.
\item We provides \emph{true uniform} distributed verification protocol by adjusting the $\mathcal{TDK}$ length to the prime or coprime to the sequence length.
\item To send the $\mathcal{TDK}$ at each SUBTPA we uses String Reconciliation Protocol based on the graph theoritic approach and reduces the usage of network bandwidth.
\item we have given analytical approach to show that each SUBTPA will detect approximately same number of errors.
\item In addition, we reduce the computational and workload at the Cloud server side by sending 10\% subsequence at a time.
\end{enumerate}
We have shown that, our protocols detect errors very fast and uses less network bandwidth and reduces the chance of network congestion.

\section{Outline of Project Report}
In chapter 2, we discusses about the Sobol Sequence and String Reconciliation protocol in details with concrete example. In chapter 3, we explain the audit system architecture and our proposed protocols and analysis for each protocols. Thereafter, in chapter 4, we analyzes performance and gives the experimental results. Finally, in chapter 6, we present a brief conclusion about our work. 

\chapter{Theoretical Underpinnings}
\label{ch:techintro}

In this chapter we reviews some theoretical underpinnings used in our protocols. First we discussed about Sobol Sequences in details and next  introduces String Reconciliation Protocols by using one concrete example.

\section{Sobol Sequence}
Sobol Sequence [4] is a low discrepancy, quasi-random sequences that generates sequences between the interval [0, 1). A salient feature of this sequence is that the sequences are uniformly distributed over the interval [0, 1) as well as segment wise with segment size $2^{i}, i\in[1,n]$. Thus, to generate a sequence of values
$x^{1}, x^{2}, x^{3}, \ldots,  0<x^{i}<1,$ with low discrepancy over the unit interval, we first need a set of direction numbers $v_{1},v_{2}, \ldots$. Each $v_{i}$  can be represented as in~(\ref{eq: sobol1}):
\begin{equation}\label{eq: sobol1}
v_{i}=m_{i}/2^{i}
\end{equation}

where $m_{i}$ is an odd integer, and $0 < m_{i} < 2^{i}$. To obtain $v_{i}$, we needs a primitive polynomial, $\mathcal{P}$, with coefficients chosen from (0, 1), over the finite field $\mathbb{Z}_{2}$, is of the form (\ref{eq: sobol2}):

\begin{equation}\label{eq: sobol2}
 X^{d} + a_{1}X^{d-1} + \cdots + a_{d-1}X + 1
\end{equation}

where each $a_{i}$ is 0 or 1 and $\mathcal{P}$ is a primitive polynomial of degree d in $\mathbb{Z}_{2}$. A Primitive Polynomial is a polynomial which is irreducible and generates all the elements of an extension field from a base field and the order of that polynomial should be $2^{d}-1$ where d is the degree of that Polynomial. As $\mathcal{P}$ is primitive,  $a_{d}$, the constant term, is necessarily equal to 1. There are $\phi(2^{d}-1)/d$ number of primitive polynomials of degree $d$, and we can choose any one randomly, but depends on the primitive polynomial the sequence will be change. Here, $\phi$ denotes the Euler function. Once we have chosen a polynomial, we use its coefficients to define a recurrence for calculating $m_{i}$ from~(\ref{eq: sobol3})

\begin{equation}\label{eq: sobol3}
m_{i} = 2a_{1}m_{i-1}\oplus 2^{2}a_{2}m_{i-2}\oplus\cdots\oplus 2^{d-1}a_{d-1}m_{i-d+1}
\oplus 2^{d}a_{d}m_{i-d}\oplus m_{i-d}
\end{equation}

The term $a_{1},a_{2},\ldots,a_{d}$ of recurrence relation~(\ref{eq: sobol3}) are the coefficients from the chosen primitive polynomial, and the values of $m_{1}, m_{2} ,\ldots,  m_{d}$ can be chosen freely provided that each $m_{i}$ is odd and $m_{i} < 2^{i}$; subsequent values $m_{d+1}$, $m_{d+2}$ ,$\ldots$ are then determined by the recurrence relation in~(\ref{eq: sobol3}). Here, also depends on the $m_{1}$, $m_{2}$ ,$\ldots$,  $m_{d}$ values the sequence will be changes.
Thus, after generating the $m_{i}$ values we can generates the $v_{i}$ values from~(\ref{eq: sobol1}).

\subsection{Sequence Generations with Different Primitive Polynomials}
For example, here we are taking, degree $d=3$ and calculates total number of primitive polynomials of degree 3 by using $\phi(2^{d}-1)/d$ and shown in~(\ref{eq: cal}).
\begin{equation}\label{eq: cal}
\begin{split}
 Total\; Number\; of\; Primitive\; Polynomial&=\phi(2^{d}-1)/d\\
                                     &=\phi(2^{3}-1)/3\\
                                     &=\phi(7)/3\\
                                     &=6/3 \\
                                     &=2 
\end{split}
\end{equation}
The Primitive Polynomials of degre 3 are in~(\ref{eq: prim1}). 
\begin{equation}\label{eq: prim1}
\begin{split}
x^{3}+x+1\\
x^{3}+x^{2}+1
\end{split}
\end{equation}  

We have shown the sequence generation for both Primitive Polynomials. First, we chooses the polynomial~(\ref{eq: sobol4}) for sequence generation.
\begin{equation}\label{eq: sobol4}
x^{3}+x+1
\end{equation}
Hence, the coefficients are $a_{1}=0,a_{2}=1$.
By applying the coefficient values in~(\ref{eq: sobol3}), we get the corresponding recurrence is in~(\ref{eq:sobol5})
\begin{equation}\label{eq:sobol5}
\begin{split}
 m_{i} &=2a_{1}m_{i-1} \oplus 2^{3-1}a_{3-1}m_{i-3+1} \oplus 2^{3}m_{i-3}\oplus m_{i-3}\\
       &=2a_{1}m_{i-1} \oplus 2^{2}a_{2}m_{i-2} \oplus 2^{3}m_{i-3} \oplus m_{i-3}\\
       &=2^{2}m_{i-1} \oplus 2^{3}m_{i-3} \oplus m_{i-3}\\
       &=4m_{i-2} \oplus 8m_{i-3} \oplus m_{i-3}
\end{split}
\end{equation}
For solving the recurrence relation~(\ref{eq:sobol5}) we chooses the initial values $m_{1} = 1$, $m_{2} = 3$, and $m_{3} = 7$, satisfy the condition $m_{i}<2^{i}$ and odd. Then, $m_{4}$, $m_{5}$,$\ldots$ are calculated from~(\ref{eq:sobol5}) as follows:: 
\begin{equation*}
 \begin{split}
 m_{4} & = 12\oplus8\oplus1\\
      & = 1100 \oplus 1000 \oplus 0001\\
      & = 0101    \\
      & = 5
 \end{split}
\end{equation*}

\begin{equation*}
 \begin{split}
  m_{5} & = 28 \oplus 24 \oplus 3 \\
        & = 11100 \oplus 11000 \oplus 00011 \\
        & = 00111     \\
        & = 7
\end{split}
\end{equation*}

\begin{equation*}
 \begin{split}
  m_{6} & = 20 \oplus 56 \oplus 7 \\
        & = 010100 \oplus 111000 \oplus 000111 \\
        & = 101011    \\
        & = 43
\end{split}
\end{equation*}

and so on. Therefore, by using $m_{i}$ values we first generates direction numbers $v_{i}$ from equation~(\ref{eq: sobol1}) and here we have shown upto $v_{5}$ in Table~\ref{table: dir1}.

\begin{table}[htbp]
\begin{center}
\footnotesize{
\begin{tabular}{|c|c|c|c|c|c|}
\hline
 i & 1 & 2 & 3 & 4 & 5 \\
\hline
 $m_{i}$ & 1 & 3 & 5 & 3 & 29 \\
 $v_{i}$ & 0.5 & 0.75 & 0.875 & 0.3125 & 0.21875 \\
 in binary & 0.1 & 0.11 & 0.111 & 0.0101 & 0.00111 \\
\hline
\end{tabular}
}
\end{center}
\caption{\normalsize Direction Number for $x^{3}+x+1$} \label{table: dir1}
\end{table} 

To generate the sequence $x^{1}, x^{2},\ldots$ we can use equation~(\ref{eq: sobol6})
\begin{equation} \label{eq: sobol6}
x^{n} = g_{1}v_{1} \oplus g_{2}v_{2} \oplus \ldots
\end{equation}
where $\ldots g_{3}g_{2}g_{1}$ is the gray code representation of n.

\textit{Property of gray code~\cite{sobol2}}:\\
The Gray code for $n$ and the Gray code for $n + 1$ differ in only one position.
If $b_{c}$ is the rightmost zero-bit in the binary representation of $n$ (add a leading
zero to n if there are no others), then $g_{c}$ is the bit whose value changes.

Now using the above property, we can generate $x^{n+1}$ in terms of $x^{n}$ as in~(\ref{eq:sobol7}).

\begin{equation}\label{eq:sobol7}
x^{n+1}=x^{n}+v_{c}
\end{equation}

Where $x^{0}=0$, and $b_{c}$, is the rightmost zero-bit in the binary representation of n.
Now by using Table~\ref{table: dir1}~\cite{sobol2} we calculates Sobol Sequence as follows:
\begin{equation*}
 \begin{split}
  x^{0} &=0\\
n &=0\\
c &=1
\end{split}
\end{equation*}
(First rightmost zero-bit in n is in $1^{th}$ position so $c=1$)
\begin{equation*}
 \begin{split}
  x^{1} &=x^{0} \oplus v_{1}\\
&=0.0 \oplus 0.1 \\
&=0.1 = \frac{1}{2} =0.5\\
n&= 1=01\\
c&=2
\end{split}
\end{equation*}

\begin{equation*}
 \begin{split}
  x^{2} &=x^{1} \oplus v_{2}\\
&=0.10 \oplus 0.11 \\
&=0.01 = \frac{1}{4} =0.25\\
n&=2=10\\
c&=1
\end{split}
\end{equation*}

\begin{equation*}
 \begin{split}
  x^{3} &=x^{2} \oplus v_{1}\\
&=0.01 \oplus 0.10 \\
&=0.11 = \frac{3}{4} =0.75\\
n&=3=011\\
c&=3
\end{split}
\end{equation*}

\begin{equation*}
 \begin{split}
  x^{4} &=x^{3} \oplus v_{3}\\
&=0.110 \oplus 0.111 \\
&=0.001 = \frac{1}{8} =0.125\\
n&=4=0100\\
c&=1
\end{split}
\end{equation*}
\begin{equation*}
 \begin{split}
  x^{5} &=x^{4} \oplus v_{1}\\
&=0.001 \oplus 0.100 \\
&=0.101 = \frac{5}{8} =0.625\\
n&=5=101\\
c&=2
\end{split}
\end{equation*}
\begin{equation*}
 \begin{split}
  x^{6} &=x^{5} \oplus v_{2}\\
&=0.101 \oplus 0.110 \\
&=0.011 = \frac{3}{8} =0.375\\
n&=6=110\\
c&=1
\end{split}
\end{equation*}
\begin{equation*}
 \begin{split}
  x^{7} &=x^{6} \oplus v_{1}\\
&=0.011 \oplus 0.100 \\
&=0.111 = \frac{7}{8} =0.875\\
n&=7=0111\\
c&=4
\end{split}
\end{equation*}
\begin{equation*}
 \begin{split}
  x^{8} &=x^{7} \oplus v_{4}\\
&=0.1110 \oplus 0.0101 \\
&=0.1011 = \frac{11}{16} =0.6875\\
n&=8=1000\\
c&=1
\end{split}
\end{equation*}
and so on. Hence, the generated  sequences are in~(\ref{eq: seq1})
\begin{equation}\label{eq: seq1}
\begin{split}
\mathcal{L}= \{0, 0.5, 0.25, 0.75, 0.125,0.625,0.375\\
0.875,0.6875,0.1875,0.9375,0.4375,0.5625,\ldots\}
\end{split}
\end{equation}
The sequence will be change depends on the particular Primitive Polynomial and on initial values of $m_{i}$, $i\in\{1,2,\ldots,d\}$. In our protocols we multiply constant powers of two to convert the sequences into integer number sequences. Here, we multiply 64 to the $\mathcal{L}$, the resulted sequences is in~(\ref{eq: seq2})
 \begin{equation}\label{eq: seq2}
\mathcal{L}= \{0, 32, 16, 48,8,40,24,56,44,12,60,28,36,\ldots\}
\end{equation} 

Now we takes the second  Primitive Polynomial from~(\ref{eq: prim1}) and generates the sequence have shown in the below.
\begin{equation}
 x^{3}+x^{2}+1
\end{equation}
Hence, the coefficients are $a_{1}=1,a_{2}=0$.
By applying the coefficient values in~(\ref{eq: sobol3}), we get the corresponding recurrence is in~(\ref{eq: prim})
 \begin{equation}\label{eq: prim}
\begin{split}
 m_{i} &=2a_{1}m_{i-1} \oplus 2^{3-1}a_{3-1}m_{i-3+1} \oplus 2^{3}m_{i-3}\oplus m_{i-3}\\
       &=2a_{1}m_{i-1} \oplus 2^{2}a_{2}m_{i-2} \oplus 2^{3}m_{i-3} \oplus m_{i-3}\\
       &=2m_{i-1} \oplus 2^{3}m_{i-3} \oplus m_{i-3}\\
       &=2m_{i-1} \oplus 8m_{i-3} \oplus m_{i-3}
\end{split}
\end{equation}

For solving the recurrence relation~(\ref{eq: prim}) we chooses the initial values $m_{1} = 1$, $m_{2} = 3$, and $m_{3} = 5$, satisfy the condition $m_{i}<2^{i}$ and odd. Then, $m_{4}$, $m_{5}$,$\ldots$ are calculated from~(\ref{eq: prim}) as follows:: 

\begin{equation*}
\begin{split}
m_{4} &=2m_{3} \oplus 8m_{1} \oplus m_{1}\\
     &= 10 \oplus 8 \oplus 1\\
     &=0011 \;(in\; binary)\\
     &=3
\end{split}
\end{equation*}
\begin{equation*}
\begin{split}
m_{5} &=2m_{4} \oplus 8m_{2} \oplus m_{2}\\
     &= 6\oplus 24 \oplus 3\\
     &=11101 \;(in\; binary)\\
     &=29
\end{split}
\end{equation*}
\begin{equation*}
\begin{split}
m_{6} &=2m_{5} \oplus 8m_{3} \oplus m_{3}\\
     &= 58\oplus 40 \oplus 5\\
     &=010111 \;(in\; binary)\\
     &=23
\end{split}
\end{equation*}  
and so on. Therefore, by using $m_{i}$ values we first generates direction numbers $v_{i}$ from equation~(\ref{eq: sobol1}) and here we have shown upto $v_{5}$ in Table~\ref{table: dir2}. 

\begin{table}[htbp]
\begin{center}
\footnotesize{
\begin{tabular}{|c|c|c|c|c|c|}
\hline
 i & 1 & 2 & 3 & 4 & 5 \\
\hline
 $m_{i}$ & 1 & 3 & 5 & 3 & 29 \\
 $v_{i}$ & 0.5 & 0.75 & 0.625 & 0.1875 & 0.90625 \\
 in binary & 0.1 & 0.11 & 0.101 & 0.0011 & 0.11101 \\
\hline
\end{tabular}
}
\end{center}
\caption{\normalsize Direction Number for $x^{3}+x^{2}+1$} \label{table: dir2}
\end{table} 

The sequence can be generated as follows:\\
\begin{equation*}
 \begin{split}
  x^{0} &=0\\
n &=0\\
c &=1
\end{split}
\end{equation*}

\begin{equation*}
 \begin{split}
  x^{1} &=x^{0} \oplus v_{1}\\
&=0.0 \oplus 0.1 \\
&=0.1 = \frac{1}{2} =0.5\\
n&= 1=01\\
c&=2
\end{split}
\end{equation*}

\begin{equation*}
 \begin{split}
  x^{2} &=x^{1} \oplus v_{2}\\
&=0.10 \oplus 0.11 \\
&=0.01 = \frac{1}{4} =0.25\\
n&=2=10\\
c&=1
\end{split}
\end{equation*}

\begin{equation*}
 \begin{split}
  x^{3} &=x^{2} \oplus v_{1}\\
&=0.01 \oplus 0.10 \\
&=0.11 = \frac{3}{4} =0.75\\
n&=3=011\\
c&=3
\end{split}
\end{equation*}

\begin{equation*}
 \begin{split}
  x^{4} &=x^{3} \oplus v_{3}\\
&=0.110 \oplus 0.101 \\
&=0.011 = \frac{3}{8} =0.375\\
n&=4=100\\
c&=1
\end{split}
\end{equation*}

\begin{equation*}
 \begin{split}
  x^{5} &=x^{4} \oplus v_{1}\\
&=0.011 \oplus 0.100 \\
&=0.111 = \frac{7}{8} =0.875\\
n&=5=101\\
c&=2
\end{split}
\end{equation*}

\begin{equation*}
 \begin{split}
  x^{6} &=x^{5} \oplus v_{2}\\
&=0.111 \oplus 0.110 \\
&=0.001 = \frac{1}{8} =0.125\\
n&=6=110\\
c&=1
\end{split}
\end{equation*}

\begin{equation*}
 \begin{split}
  x^{7} &=x^{6} \oplus v_{1}\\
&=0.001 \oplus 0.100 \\
&=0.101 = \frac{5}{8} =0.625\\
n&=7=0111\\
c&=4
\end{split}
\end{equation*}
\begin{equation*}
 \begin{split}
  x^{8} &=x^{7} \oplus v_{4}\\
&=0.1010 \oplus 0.0011 \\
&=0.1001 = \frac{9}{16} =0.5625\\
n&=8=1000\\
c&=1
\end{split}
\end{equation*}
and so on upto sequence length. Hence, the generated  sequences are in~(\ref{eq: seq3})

\begin{equation}\label{eq: seq3}
\begin{split}
\mathcal{L}= \{0, 0.5, 0.25, 0.75, 0.375,0.875,0.125,0.625,\\0.5625,0.0625,0.8125,0.3125,0.9375,\ldots\}
\end{split}
\end{equation}
Here, we multiply 64 to the $\mathcal{L}$, the resulted sequences is in~(\ref{eq: seq4})
 \begin{equation}\label{eq: seq4}
\mathcal{L}= \{0,32,16,48,24,56,8,40,36,4,52,20,60\ldots\}
\end{equation} 

Now we change the initial values to $m_{1}=1$, $m_{2}=3$ and $m_{3}=7$ and apply the same procedure then we shall get the sequence for Primitive Polynomial $x^{3}+x^{2}+1$ is in~(\ref{eq: seq5})
\begin{equation}\label{eq: seq5}
\begin{split}
\mathcal{L}= \{0, 0.5, 0.25, 0.75,0.125,0.625,0.375,\\0.875,0.5625,0.0625,0.8125,0.3125,0.6875,\ldots\}
\end{split}
\end{equation}
Here, we multiply 64 to the $\mathcal{L}$, the resulted sequences is in~(\ref{eq: seq6})
 \begin{equation}\label{eq: seq6}
\mathcal{L}= \{0,32,16,48,8,40,24,56,36,4,52,20,44,\ldots\}
\end{equation} 
As the sequence length will be increase the sequence elemens should be random. Therefore, by choosing polynomial we can change the sequence or by choosing initial values we can change the sequence.
\section{String Reconciliation}
The general String reconciliation protocol, Sachin et al. ~\cite{strrecon} states that if two distinct hosts \emph{A} and \emph{B}, holding two string $\sigma_{A}$ and $\sigma_{B}$, then by applying this protocol host \emph{A} will know $\sigma_{B}$ and host \emph{B} will know $\sigma_{A}$ with \emph{minimum communication and optimal computational complexity}. In this protocol hosts are independently divides their strings into multiset of ``puzzle pieces'' by using predetermined mask length and multisets are stored using Modified de Bruijn graph, and enumerates the \emph{Eulerian cycles} to get the index of the original string. The  multisets are reconciled using Set Reconciliation Protocol~\cite{setrecon}. At the final step, each host independently construct another's modified de Bruijn graph from reconciled multisets and enumerates the \emph{Eulerian Cycles} and use the index given by other's to decides the other host's string data.

\begin{figure}[tbhp]
\centering
\includegraphics[scale=0.7]{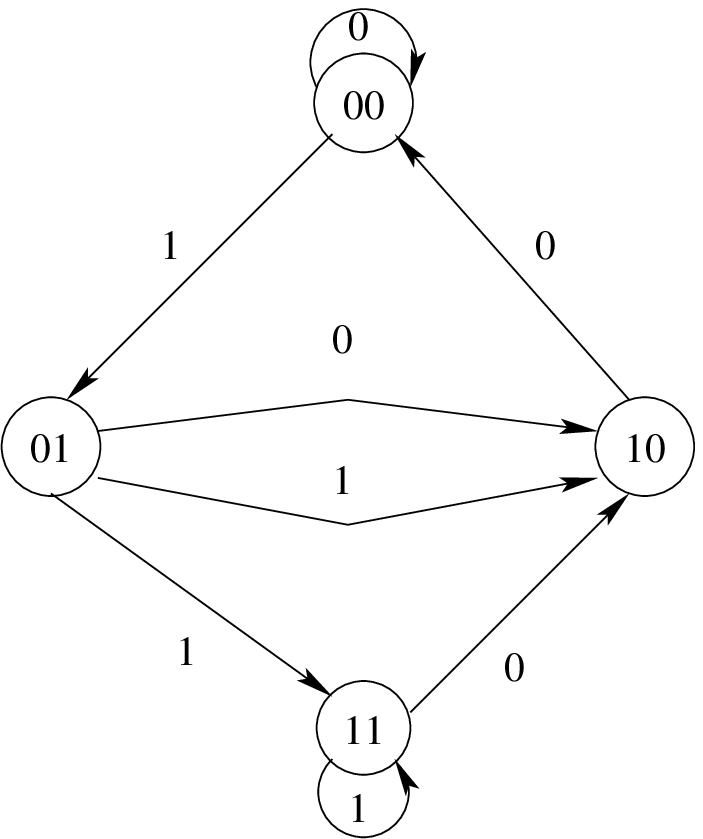}
\caption{de Bruijn Graph}
\label{fig: 1}
\end{figure}

\begin{figure}[tbhp]
\centering
\includegraphics[scale=0.7]{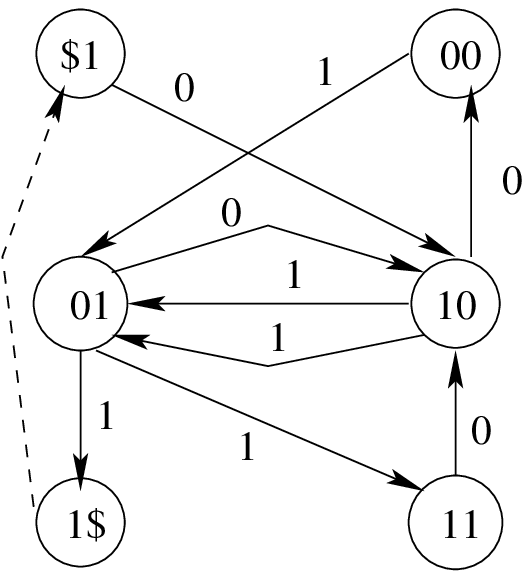}
\caption{Modified de Bruijn Graph for \$101101001\$}
\label{fig: md}
\end{figure}

The \emph{de Bruijn digraph} $G_{l_{m}}(\sum)$ over an alphabet $\sum$ and length $l_{m}$ is defined to contain $|\sum|^{l_{m}-1}$ vertices, each corresponding to a length $l_{m}-1$ string over the alphabet. There will be an edge from vertex $X$ to $Y$ exists with label $l$ if the string associated with $Y$ contains the last $l_{m}-2$ characters of $X$ followed by $l$. Thus, each edge $(X, Y)$ represents a length $l_{m}$ string over $\sum$. An example of the de Bruijn digraph $G_{3}(\{0, 1\})$, from~\cite{strrecon} , is shown in figure~\ref{fig: 1}.
In de Bruijn graph number of Eulerian cycle if fixed and it is one but in String Reconciliation protocol, number of Eulerian cycles are huge due to multiple edges in the modified de Bruijn graph. Next, we restates the steps needed to convert a de Bruijn graph to a modified de Bruijn graph.  
\begin{enumerate}
\item Parallel edges are added to the digraph for each occurrence of the same puzzle pieces.
\item Edges which represent strings not in the multiset are deleted.
\item Vertices with degree zero are deleted.
\item Two new vertices and edges corresponding to the first and last pieces of the encoded string are added.
\item An artificial edge is added between the two new vertices to make their in-degree equal their outdegree
\end{enumerate}
From the modified debruijn graph we can see that there is a directed edge between $X=x_{1}x_{2}x_{3}\ldots x_{l_{m}-1}$ and $Y=y_{1}y_{2}\ldots y_{l_{m}-1}$ iff $x_{i}=y_{i-1}$ for $i=2,3,\ldots,l_{m}-1$. So, each edge is labeled by a string of length $l_{m}$, namely $xy_{l_{m}-1}$. For example, we have given a string \$101101001\$, and corresponding multiset \{ \$10, 101, 011, 110, 101, 010, 100, 001, 01\$ \}, and the Modified de Bruijn graph is in Figure~\ref{fig: md}.

Now, consider a connected~\cite{prot}, directed graph made of a certain number of labeled nodes. A node $i$ may be connected to a node $j$ by a directed arc. If from a starting node $v_{0}$ one may go through a collection of arcs to reach an ending node $v_{m}$ in such a way that each arc is passed only once, then it is called an \emph{Eulerian path}. If $v_{0}$ and $v_{m}$ coincide the path becomes an \emph{Eulerian loop}. A graph in which there exists an Eulerian loop is called an \emph{Eulerian graph}. An Eulerian path may be made an Eulerian loop by drawing an auxiliary arc from $v_{m}$ back to $v_{0}$ .
 
From a node there may be $d_{out}$ arcs going out to other nodes, $d_{out}$ is called the outdegree of the node. There may be $d_{in}$ arcs coming into a node, $d_{in}$ being the indegree of the node. The condition for a graph to be Eulerian was indicated by Euler in 1736 and consists in
\[d_{in}(i) = d_{out}(i) \equiv d_{i} = an\; even\; number\]

for all nodes i. Numbering the nodes in a certain way, we may put their indegrees as a diagonal matrix:
    \[M = diag(d_{1} , d_{2} ,\ldots, d_{m} )\]
The connectivity of the nodes may be described by an adjacent matrix A = \{$a_{ij}$ \}, where $a_{ij}$ is the number of arcs leading from node $i$ to node $j$.

From the M and A matrices one can forms the Kirchhoff matrix:
                       \[C = M - A\]
The Kirchhoff matrix has the salient feature is that, sum of its elements along any row or column is zero. Further more, for an $m \times m$ Kirchhoff matrix all $(m - 1) \times (m - 1)$ minors are equal and we denote it by $\Delta$.

A graph is called simple if between any pairs of nodes there are no parallel arcs and there
are no rings, i.e., $a_{ij}$= 0 or 1 $\forall i,j$ and $a_{ii}$ = 0 $\forall i$. The number R of Eulerian loops in a simple Euler graph is given by

\textbf{The BEST Theorem~\cite{Euler}:} (BEST stands for N. G. de \textbf{B}ruijn, T. van Aardenne-\textbf{E}hrenfest, C. A. B. \textbf{S}mith, and W. T. \textbf{T}utte):
\begin{equation}\label{eq: best1}
 R= \Delta \prod_{i}(d_{i}-1)! 
\end{equation}

For general Euler graphs, however, there may be arcs going out and coming into one and the same node (some $a_{ii}\ne 0$) as well as parallel arcs leading from node $i$ to $j$ ($a_{ij} > 1$). It is enough to put auxiliary nodes on each parallel arc and ring to make the graph simple. The derivation goes just as for simple graphs and the final result is one has the original graph without auxiliary nodes but with $a_{ii} = 0$ and $a_{ij} > 1$ incorporated into the adjacent matrix A. However, in accordance with the unlabeled nature of the parallel arcs and rings one must eliminate the redundancy in the counting result by dividing it by $a_{ij} !$. Thus the BEST formula is modified~\cite{prot} to
\begin{equation}\label{eq: best2}
 R= \frac{\Delta \prod_{i}(d_{i}-1)!}{\prod_{ij}a_{ij}!}
\end{equation}

As $0! = 1!$ = 1 Eq.~(\ref{eq: best1}) reduces to (\ref{eq: best2}) for simple graphs.

Here we restate the String Reconciliation protocol for the completeness of our discussion:\\

\textbf{STRING-RECON}\\
Two hosts, A and B, holding strings $\sigma_{A}$ and $\sigma_{B}$, respectively. The mask length is predetermined by the two hosts. Host A determines $\sigma_{B}$ and host B determines $\sigma_{A}$ as follows:
\begin{enumerate}
\item[1.] Host A and B independently transforms their strings, $\sigma_{A}$ and $\sigma_{B}$ into multiset of pieces $MS_{A}$, $MS_{B}$ and constructs modified de Bruijn digraph $dG_{A}$, $dG_{B}$ respectively.
\item[2.] Host A enumerates all Eulerian Cycles by using backtracking algorithm on $dG_{A}$ to determine the index,$n_{A}$, corresponding to $\sigma_{A}$, B also perform same task on $dG_{B}$ and decides index, $n_{B}$, corresponding to $\sigma_{B}$.
\item[3.] A and B transform multisets $MS_{A}$ and $MS_{B}$ into sets, $S_{A}$ and $S_{B}$ by concatenating each element in the multiset with the number of times it occurs and hashing the result.
\item[4.] The CPIsync algorithm is executed to reconcile the sets $S_{A}$ and $S_{B}$. In addition, A sends $n_{A}$ to B and B
sends $n_{B}$ to A. At the end of this step, both A and B know $S_{A}, S_{B}, n_{A}$, and $n_{B}$. Afterthat, host A sends
multiset elements corresponding to the set $S_{A}\backslash S_{B}$ to B and B sends multiset elements corresponding to the set $S_{B} \backslash S_{A}$ to A. Thus A has $MS_{A}, MS_{B}$ and B has $MS_{B}, MS_{A}$.
\item[5.] Now A and B construct modified de Bruijn digraphs from $MS_{B}$ and $MS_{A}$, respectively.
\item[6.] Host A enumerates Eulerian Cycles on $MS_{B}$, and use the index $n_{B}$ to determine $\sigma_{B}$, similarly B also perform the same task and determine the string $\sigma_{A}$.
\end{enumerate}
Now we are going to describes the String Reconciliation Protocol by using one concrete example, where host A and B holds two strings \$10010101\$ and \$101101001\$ respectively. All the below steps are performed by host A and B simultaneously.
\begin{enumerate}
 \item[{\sl Step1:}] Host A holds the string \[\sigma_{A} = 10010101\] and host B holds the string \[\sigma_{B}= 101101001\]

\item[{\sl Step2:}] To indicate the starting and ending point of the string both host will add a \emph{special character not present in the original string} at the both end of the strings. Let the spccial character is \$.
\[\sigma_{A} = \$10010101\$\] and \[\sigma_{B}= \$101101001\$\]

\item[{\sl Step3:}] Depending on the predetermined mask length both host A and B independently creates the multisets as follows and shown in~(\ref{eq:multisetA}).\\
\$10010101\$\\
 111\\
\$10010101\$\\
\hspace*{0.19cm}111\\
\$10010101\$\\
\hspace*{0.36cm}111\\
\$10010101\$\\
\hspace*{0.45cm} 111\\
\$10010101\$\\
\hspace*{0.6cm} 111\\
\$10010101\$\\
 \hspace*{0.79cm} 111\\
\$10010101\$\\
 \hspace*{0.97cm} 111\\
\$10010101\$\\
\hspace*{1.1cm}  111\\
\begin{equation}\label{eq:multisetA}
MS_{A}=\{ \$10, 100, 001, 010, 101, 010, 101, 01\$ \}
\end{equation}
We apply the similar procedure to creates the multiset for host B and shown in~(\ref{eq:multisetB})

\begin{equation}\label{eq:multisetB}
MS_{B}=\{ \$10, 101, 011, 110, 101, 010, 100, 001, 01\$ \}
\end{equation}

\item[{\sl Step4:}] From the multiset $MS_{A}$ and $MS_{B}$ both host independently construct Modified de Bruijn graph where each edge represents one multiset element. The graphs has been shown in figure~\ref{fig:2} for host A and in figure~\ref{fig:3} for host B respectively.

\begin{figure}[thbp]
\centering
\subfigure[Modified de Bruijn Graph for Host A]{
   \includegraphics[scale=0.6]{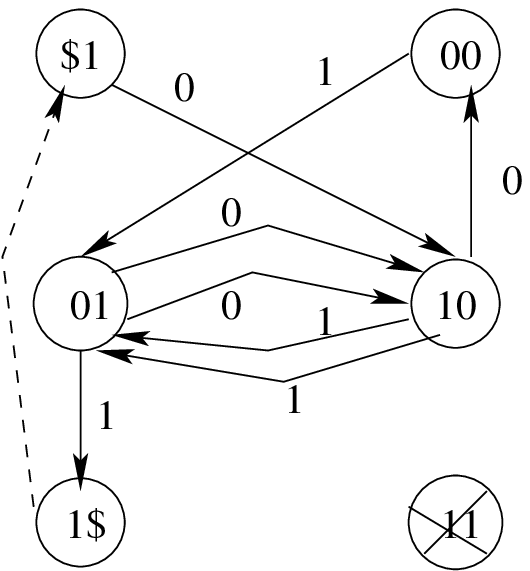}
   \label{fig:2}
 }
 \subfigure[Modified de Bruijn Graph for Host B]{
   \includegraphics[scale=0.6]{debruijn2.eps}
   \label{fig:3}
 }
\caption{Modified de Bruijn Graph}
\label{fig: deb}
\end{figure}

\item[{\sl Step5:}]Now, host A and B independently enumerates all the Eulerian Cycle by using \emph{Backtracking Algorithm} and find the index of the particular Euler cycle that gives the corresponding string. Let, $n_{A}$ is the index at host A that gives the string $\sigma_{A}$ and $n_{B}$ is the index at host B that gives the string $\sigma_{B}$ respectively.\\
\hspace*{1cm} \textbf{Host A}  \hspace{1.8in}             \textbf{Host B}\\
Cycle:1: \$10101001\$          \hspace{1in}         Cycle:1: \$101101001\$ \\
Cycle:2: \$10100101\$          \hspace{1in}         Cycle:2: \$101100101\$ \\
Cycle:3: \$10101001\$          \hspace{1in}         Cycle:3: \$101011001\$\\
Cycle:4: \$10100101\$          \hspace{1in}         Cycle:4: \$101001101\$ \\
Cycle:5: \$10101001\$          \hspace{1in}         Cycle:5: \$101101001\$ \\ 
Cycle:6: \$10100101\$          \hspace{1in}         Cycle:6: \$101100101\$\\
Cycle:7: \$10101001\$          \hspace{1in}         Cycle:7: \$101011001\$\\
Cycle:8: \$10100101\$          \hspace{1in}         Cycle:8: \$101001101\$\\
Cycle:9: \$10010101\$          \hspace{1in}         Cycle:9: \$100110101\$\\
Cycle:10: \$10010101\$         \hspace{0.9in}       Cycle:10: \$100110101\$\\
Cycle:11: \$10010101\$         \hspace{0.9in}       Cycle:11: \$100101101\$\\
Cycle:12: \$10010101\$         \hspace{0.9in}       Cycle:12: \$100101101\$\\
Thus we can take $n_{A}$ equal to any one from 9-12 and $n_{B}$ equal to 1 or 5.

\item[{\sl Step6:}] Both host A and B transform Multiset into set as follows:\\
$MS_{A}=\{ \$10, 100, 001, 010, 101, 010, 101, 01\$ \}$\\
Let, \$ is replaced by $1000101$\\
{\small $MS_{A}=\{ \underbrace{1000101}_{\$}1001, 10001, 00101, 01010, 10110, 010, 101, 01\underbrace{1000101}_{\$}01 \}$}\\
corresponding Decimal values in set are\\
$S_{A}= \{1113,17,5,10,22,789\}$\\
Here, we use modulo 47 hashing function to hash the result:\\
$S_{A}= \{32,17,5,10,22,37\}$

Similarly perform the above steps at Host B:\\
$MS_{B}=\{ \$10, 101, 011, 110, 101, 010, 100, 001, 01\$ \}$\\
{\small $MS_{B}=\{ \underbrace{1000101}_{\$}1001, 10110, 01101, 11001, 01001, 10001, 00101, 01\underbrace{1000101}_{\$}01 \}$}\\
$S_{B}= \{1113,22,13,25,9,17,5,789\}$\\
Here, we use modulo 47 hashing function to hash the result:\\
$S_{A}= \{32,22,13,25,9,17,5,37\}$

\item[{\sl Step7:}] Now, we uses \emph{Set Reconciliation Protocol}~\cite{setrecon} to reconcile the sets $S_{A}$ and $S_{B}$ so that both host will get $S_{A}\cup S_{B}$. According to Set reconciliation protocol, we need a upper bound of the difference between two sets, $\overline{m}$. Let, $\overline{m}$=5 and we have to take 5 sample point that are not present in $S_{A}$ and $S_{B}$.\\
Let, $E=\{-1,-2,-3,-4,-5\}$\\
Host A and B converts set to Characteristic Polynomial whose roots denotes the set elements and calculates for $Z\in \{-1,-2,-3,-4,-5\}$ and the final result is in~(\ref{eq:paira}) and ~(\ref{eq:pairb}) \\
$\chi_{S_{A}}(Z)=(Z-32)(Z-17)(Z-5)(Z-10)(Z-22)(Z-37)$\\
$\chi_{S_{B}}(Z)=(Z-32)(Z-22)(Z-13)(Z-25)(Z-9)(Z-17)(Z-5)(Z-37)$\\
Here we calculate all the calculation over the Finite Field $F_{q}$ where $q$ satisfied the equation~(\ref{eq:flimit})
\begin{equation}\label{eq:flimit}
 q > 2^{b} + \overline{m}
\end{equation}
Where b is the maximum length of the set element in Binary notation.

Here, We take q=83.
\begin{equation}\label{eq:paira}
 \{(-1,70)(-2,1)(-3,36)(-4,32)(-5,53)\}
\end{equation}

\begin{equation}\label{eq:pairb}
 \{(-1,67)(-2,60)(-3,24)(-4,53)(-5,69)\}
\end{equation}

\item[{\sl Step8:}] Now host A send $|S_{A}|$=6 and the value pair is in~(\ref{eq:paira}) to B.Thus B has\\
\{(-1,70)(-2,1)(-3,36)(-4,32)(-5,53)\} and $|S_{A}|$ =6 and\\
\{(-1,67)(-2,60)(-3,24)(-4,53)(-5,69)\} and $|S_{B}|$=8.\\
Now, B calculates Rational Function, \\
$f(Z)= \chi_{S_{A}}(Z)/ \chi_{S_{B}}(Z)$ for $Z\in \{-1,-2,-3,-4,-5\}$ as follows::\\
$f(-1)=\frac{70}{67}\quad mod\; 83 =6$\\
$f(-2)=\frac{1}{60}\quad mod\; 83 =18$\\
$f(-3)=\frac{36}{24}\quad mod\; 83 =43$\\
$f(-2)=\frac{32}{53}\quad mod\; 83 =10$\\
$f(-2)=\frac{53}{69}\quad mod\; 83 =14$

Let, represent the points and the corresponding rational function values in set $V$ as follows:\\
 $V=\{(-1,6)(-2,18)(-3,43)(-4,10)(-5,14)\}$.\\
Now we have to Interpolate these values so that Rational Function can be generated that gives the difference between sets $S_{A}$ and $S_{B}$.

We can calculate the degree of the numerator and denominator of the Rational function as follows::\\
Let, $d=|S_{A}|-|S_{B}|= 6-8 -2$.\\
Now let, $m_{A}$ and $m_{B}$ are the degree of the numerator and denominator respectively and can be calculated as follows:\\
\begin{equation*}
 \begin{split}
  m_{A}&\leqslant \lfloor (\overline{m}+d)/2 \rfloor \\
  & \leqslant \lfloor (5-2)/2 \rfloor\\
& \leqslant \lfloor 3/2 \rfloor\\
& \leqslant 1
\end{split}
\end{equation*}

\begin{equation*}
 \begin{split}
  m_{B} & \leqslant \lfloor (\overline{m}-d)/2 \rfloor \\
  & \leqslant \lfloor (5-(-2))/2 \rfloor\\
& \leqslant \lfloor 7/2 \rfloor\\
& \leqslant 3
\end{split}
\end{equation*}
Therefore the highest degree of numerator is 1 and denominator is 3. Here, the polynomial should be \emph{monic}. Hence, heighest degree coefficient must be 1. Therefore, Rational function is of the form in~(\ref{eq:rf1}).
\begin{equation}\label{eq:rf1}
\begin{split}
f_{i}=\frac{x_{i}+p_{0}}{x^{3}_{i}+q_{2}x^{2}_{i}+q_{1}x_{i}+q_{0}}\\
x_{i}+p_{0}=f_{i}(x^{3}_{i}+q_{2}x^{2}_{i}+q_{1}x_{i}+q_{0})
\end{split}
\end{equation}
Now we uses Gauss Elimination Method to Interpolate the Rational function and we shall get the difference between the set as in~(\ref{eq:rf2})
\begin{equation}\label{eq:rf2}
\begin{split}
f_(x) &=\frac{x_{i}+73}{x^{3}+36x^{2}+3x+63}\\
      &= \frac{(x-10)}{(x-25)(x-9)(x-13)} \\
      &= \frac{S_{A}\backslash S_{B}}{S_{B}\backslash S_{A}}
\end{split}
\end{equation}
In~(\ref{eq:rf2}) we factorize (discussed in Appendix A)  the rational function and gets the roots, difference between two sets. So, \{10\} is not present in $B's$ set but present in $A's$ set similarly \{25,9,13\} are not present in $A's$ set but is in $B's$ set. The above Rational function interpolation is performed by host A also. 

\item[{\sl Step9:}] \textbf{B construct $A's$ set as follows}\\
B has, $S_{A}\backslash S_{B}=\{10\}$\\
       $S_{B}\backslash S_{A}=\{25,9,13\}$\\

\begin{equation*}
\begin{split}
S_{B}'&=S_{B}\backslash (S_{B}\backslash S_{A}) \\
      &=\{32,17,5,22,37,25,9,13\}\backslash \{25,9,13\}\\
      &=\{32,17,5,22,37\}\\
S_{B}''&=S_{B}'\cup (S_{A}\backslash S_{B}) \\
      &=\{32,17,5,22,37\}\cup \{10\}\\
      &=\{32,17,5,22,37,10\}\\
      &=S_{A} 
\end{split}
\end{equation*}

So, Now B has $S_{A}$ and $S_{B}$. Similarly A also performed the same steps and will get $S_{B}$.

\item[{\sl Step10:}] Now host A sends the multiset element corresponding to $S_{A}\backslash S_{B}$ to B.
\begin{equation*}
\begin{split}
MS_{A}=\{ \$10, 100, 001, \underline{010}, 101, \underline{010}, 101, 01\$ \}\\
S_{A}= \{32,17,5,\underline{10},22,37\}
\end{split}
 \end{equation*}

$S_{A}\backslash S_{B}=\{10\}$ and the corresponding multiset element is \{010\} and it occurs two times in $MS_{A}$. Hence, A sends \{010,2\} and index $n_{A}=9$ to B. 

Now B construct $A's$ multiset as follows:
\begin{equation}
\begin{split}
MS_{B}'&=\{\$10, 101, 011, 110, 101, 010, 100, 001, 01\$ \} \backslash \{110,010,011\}\\ 
       &= \{ \$10,101,101,100,001,01\$\}\\
MS_{B}'' &=MS_{B}'\cup \{010,010\}\\
         &= \{ \$10,101,101,100,001,01\$\} \cup \{010,010\}\\
         &= \{ \$10,101,101,100,001,01\$,010,010\}\\
         &=MS_{A} 
\end{split}
\end{equation}

Now B has $MS_{B}$, $n_{B}$, $MS_{A}$, and $n_{A}$. Similarly A also performed these steps and get $MS_{B}$, $n_{B}$.

\item[{\sl Step11:}] Now, B construct Modified de Bruijn graph for the multiset $MS_{A}$ and use Backtraching algorithm to enumerates the Eulerian Cycle and use the index $n_{A}$ to get the string $\sigma_{A}= \$10010101\$ $. A also construct the Modified de Bruijn graph for $MS_{B}$ and enumerates all Eulerian cycle and use the index $n_{B}$ to get the string $\sigma_{B}=\$101101001\$ $.
\end{enumerate}

In this general String Reconciliation Protocol both host performed all Rational function interpolation, and factorize to get the difference between sets. It is very difficult to  But in our Distributed verification protocol Main TPA and SUBTPA will reconcile $\mathcal{TDK}$, $\sigma_{S_{i}}$ and Sobol Random Key $\sigma_{r}$, where Main TPA will know the $\sigma_{r}$ as well as all $\mathcal{TDK}$s and each SUBTPA will know only the Sobol Random Key, $\sigma_{r}$. Therefore, Main TPA have no much overhead about calculations, the rational function interpolation and factorization is performed by every SUBTPA independently and acquire the respective $\mathcal{TDK}$ assigned for it.

\section{Related Works}
Recently, many researchers have focused on the problem of remote data security and proposed different protocols [16-23] to verify the integrity of remote data stored in distributed storage systems without having a local copy of data based on Remote Data Checking (RDC) protocol.  In RDC protocol, the client generates some metadata.  Later, the verifier stores the pre-computed values and sends challenge to the server through Challenge-Response Protocol. Upon receiving a request from the verifier, the server computes a response and returns to the verifier. Then, the verifier checks whether data has been modified or not. These RDC protocols can be classified into two categories: Deterministic verification protocols and Probabilistic verification protocols [16-21]. Deterministic verification schemes give deterministic guarantee of the data integrity, because they check the integrity of all the data blocks. Deswarte et al.~\cite{deswarte} and Filho et al.~\cite{filho}  first proposed the remote data checking protocol on untrusted servers. Sebe et al.~\cite{sebe} described a protocol to check remote file integrity in critical infrastructure, and Hao et al.~\cite{p16} proposed a Privacy-preserving Remote Data checking protocol. However, their schemes are unfeasible when file size is large. The probabilistic verification schemes give Probabilistic guarantee of the data integrity and the detection probability will be high if an attacker deletes a fraction all the blocks because the challenged blocks are randomly selected. Ateniese et al.~\cite{pdp} proposed a Provable Data Possession at untrusted servers. In their subsequent work, Ateniese et al.~\cite{epdp} described a Scalable PDP, and C. Wang et al.~\cite{wangtowards} proposed a secure and dependable protocol, Q. Wang et al.~\cite{wangpublic} described a Enabling Public Integrity Verification Protocol and Zhu et al.~\cite{audit} proposed Public Dynamic Audit Service. Although Probabilistic verification schemes achieved the integrity of remote data assurance under the different systems, but they lack in provides a strong security assurance to the clients. This is due to their verifying process using pseudorandom sequence, which does not cover the entire file while generating the integrity proof. In addition all the above prior works takes much more time to detect the error also some protocol uses Third Party Auditor for public verification. Here, we proposed distributed verification protocols where many number of SUBTPA will do the integrity verification under a Main TPA/ Client. In the next chapter, we describes our protocols in details.

\chapter{Distributed Verification Protocols}
\label{ch:method}
In this chapter, we presented our proposed distributed verification protocols for checking the integrity of the stored data in the Clouds. In section1, we have given a block diagram about our proposed audit system architecture and in the subsequence section we describes our protocols.
\section{Audit System Architecture}
In this section, we introduce the audit system architecture~\cite{wangtowards} for distributed verification protocols in figure~\ref{fig:audit} . Our architecture consists of four entities for the corresponding data storage verification scheme as follows:

\emph{ Data Owner} (DO): An entity, who has large amount of data to be stored in the Cloud, can be either large enterprise or individual users. In our context, we used client, users, or data owner interchangeably.

\emph{Cloud Server} (CS): an entity, which is maintained by cloud service provider (CSP) and provides data storage services and computational resources dynamically to the data owner.

\emph{Main Third Party Auditor} (TPA): An entity, who has expertise and capabilities to manage or monitor the remotely stored data on behalf of the data owner.

\emph{SUBTPA}: This is newly defined entity in our model. Its main purpose is to verify the outsourced data and gives the audit result to the Main TPA. Here many number of SUBTPA will works under the control of Verifier. Verifier will Distributes the verification task among various SUBTPAs, ``distributed'' over a large geographical area. 

\begin{figure}[!htbh]
\centering
\includegraphics[scale=0.6]{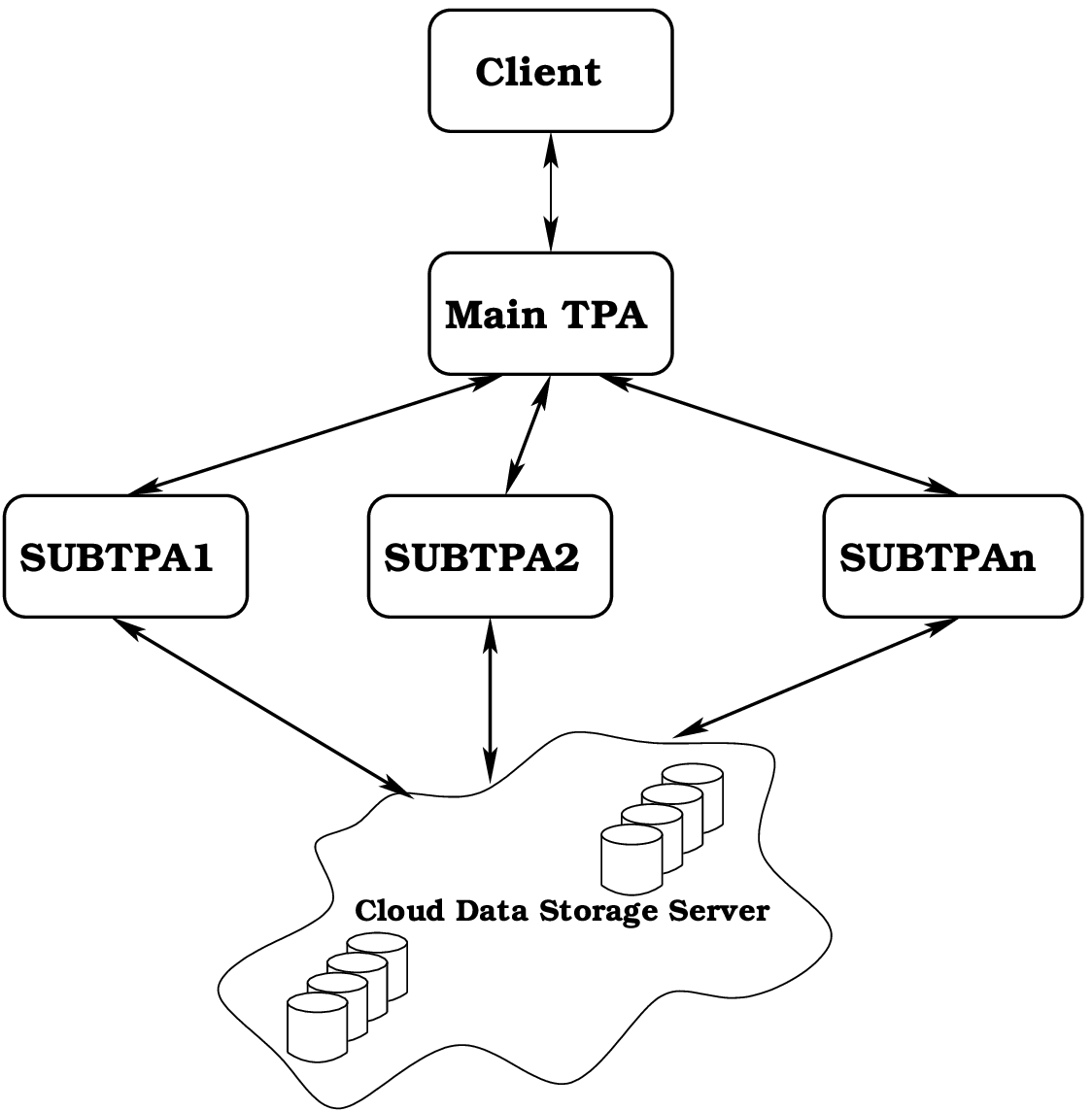}
\caption{Block Diagram of Distributed Audit System Architecture}
\label{fig:audit}
\end{figure}

Our distributed verification protocols are based on the probabilistic verification scheme and we categories our algorithms into four different cases depends on the distribution of task and challenge generation at the SUBTPA side. First, we are describing our basic protocols depends on the simple partition based approach. All the remaining protocols,  uses \emph{Task Distribution Key} to partition the task. In our protocols, we uses Sobol Random Sequences to generate the random file block numbers instead of Pseudo Random Sequences. 

To enhance the performance of our protocols, we uses (m, n) threshold scheme~\cite{threshold} with $m<n$, where Coordinator can stop the audit operation or detect the fault region after taking responses from any subset of $m$  out of $n$ SUBTPAs, because each subtask is uniformly distributed over the entire file blocks due to the use of Sobol Sequence. Here we are describing some possibilities for choosing SUBTPAs by the Coordinator as follows::

\begin{enumerate}
\item[1.] Choose SUBTPAs those are not committed in another verification work.
\item[2.] Choose SUBTPAs those has good performance in the previous time.
\item[3.] Choose SUBTPAs those are located in large geographical distant.
\item[4.] If owner know the approximate location of the storage Server, then it is better to choose SUBTPAs near Cloud organization to reduce the communication cost.
\end{enumerate}

Therefore, by using these above rule we can choose SUBTPAs for our protocols. In the next section, we describes our first protocol based on simple partition.

\section{Protocol 1: Simple Partition with Threshold Scheme}
In the first protocol, Coordinator randomly chooses one Sobol random key $\sigma_{r}$, generates the Sobol Random Block Sequences by using $f_{\sigma_{r}}(\cdot)$, where $\sigma_{r}$ consists one randomly chosen primitive polynomial, $\mathcal{P}$, of order $d$ out of $\phi(2^d-1)/d$ primitive polynomials~\cite{sobol1}, randomly chosen initial values  $m_{i}$, where $i\in\{1,2,\ldots,d\}$, $SKIP$ and $LEAP$ values respectively. In the next step, partition the generated Sequence $\mathcal{L}$ by using partition function $f_{partition}(\cdot)$, with partition length $PLen_{i}$ and denotes each subsequence as $Sub_{i}$, should maintain the equivalence relation property and also maintain the uniformity property. After generating the subsequence, $Sub_{i}$, Coordinator will distribures among various $SUBTPA_{i}$. Algorithm 1 gives the details of key generation and Distribution phase.

In Algorithm 2, we describes the distributed challenge and verification phase, where each $SUBTPA_{i}$ will independently communicates to the Cloud Servers for proof. Here, $SUBTPA_{i}$ at a time send 10\% of the subsequences to the Cloud Server as a challenge, instead of sending the whole subsequences. Therefore, it reduces the workload at the Server side as well as reduces network congestion. After sending each 10\% challenge, $SUBTPA_{i}$ will waits for the proof from \emph{any Server} because Cloud Computing is based on \emph{Distributed Control and Distributed Data} paradigm. If the  proof is matches with the stored metadata then store TRUE in its own table, Report, and send next 10\% subsequences, and waits for the next proof and this process will continue for the whole subsequence. If any mismatch will occur during proof verification, then $SUBTPA_{i}$ will immediately send a signal to the Coordinator for fault region and store FALSE in its Report table.

{\small
\begin{algorithm}[!htb]
\label{algo:Protocol1}
\SetCommentSty{sf}
\caption{Key Generation and Distribution for Protocol1}

\Indp 
Coordinator randomly chooses one Primitive Polynomial, $\mathcal{P}$ of degree $d$ and initialization number $m_{i}$, $i\in\{1,2,\ldots,d\}$ \; 
Coordinator decides Sobol Random Key as $\sigma_{r}=\langle\mathcal{P},m_{i},SKIP,LEAP,SeqLen,CONSTANT\rangle$ \\
Generates Sequences
$$\mathcal{L} \leftarrow f_{\sigma_{r}}(SeqLen)$$\\
Multiply CONSTANT \emph{powers of 2} with $\mathcal{L}$, to make each element as integer block number\;
Coordinator Determines the Number of SUBTPAs, $n$, and threshold value, $m$ \;
\For{$i \leftarrow 1$ to $n$}
{
  $Sub_{i} \leftarrow f_{partition}(PLen, \mathcal{L})$
} 
where PLen is the length of each partition, and $\mathcal{L}\subseteq \bigcup_{i\in \{1,\ldots,n\}} Sub_{i}$, 
$Sub_{\alpha}\bigcap Sub_{\beta} = \phi ~for~any~ \alpha, \beta \in \{1, \ldots, n\}$ \;
Distributes Subtask, $Sub_{i}$ to  $SUBTPA_{i}$ \;

\Indm
\end{algorithm}

\begin{algorithm}[ bht] 
\label{proc:chal1}
\caption{DistributedChalandProofVerification for Protocol1}
\Indp
$SUBTPA_{i}$ Calculates 10\% of $Sub_{i}$\;
\For {eaeh $SUBTPA_{i}$}
{
   $l \leftarrow length(Sub_{i})$\;
   $Counter_{i} \leftarrow \lfloor (10/100)*l\rfloor$\;
}

\For {$k \leftarrow 1$ to $10$ }
 {
   \For {$S \leftarrow 1$ to $Counter_{i}$ and $t\leqslant l$}
   {
      $Chal_{i,k}[s] \leftarrow Sub_{i}[t]$\;
    }
     Send $\langle Chal_{i,k}\rangle$ as a challenge to the \emph{Cloud Server}\;
      Wait for the \emph{Proof}, $PR_{i,k}$ from any Server\;
    $PR_{i,k} \leftarrow Receive()$\;
     \eIf{$PR_{i,k}$ equals to Stored Metadata}
      {
         $Report[k] \leftarrow TRUE$\;
       }
       {
         $Report[k] \leftarrow FALSE$\;
         Send Signal, $\langle Packet_{i,k},FALSE\rangle$ to the Coordinator\;
       }
}
\Indm
\end{algorithm}
}

\subsection{Analysis of Protocol 1}

\begin{figure}[!htb]
\centering
\includegraphics[scale=0.5]{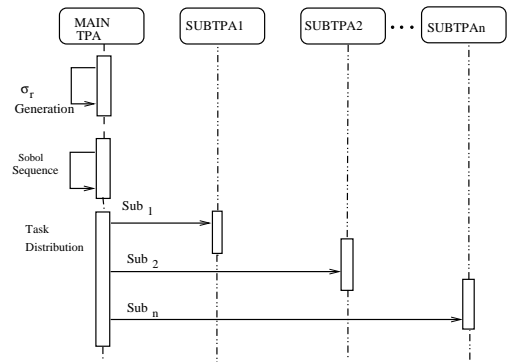}
\caption{Protocol1: Key Generation and Distribution }
\label{fig:tran1}
\end{figure}

\begin{figure}[!htb]
\centering
\includegraphics[scale=0.5]{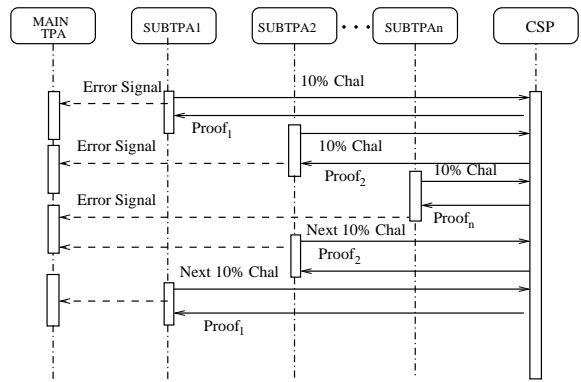}
\caption{Protocol1: Challenge and Proof Verification }
\label{fig:tran2}
\end{figure}

Protocol 1 follows the \emph{Centrally Controlled and Distributed Data} paradigm, where all SUBTPAs are controlled by the Coordinator but communicates to any Cloud Data Storage Server for verification. Here, Coordinator will decides the partition length, $PLen_{i}$, and divides the sequence among each $SUBTPA_{i}$. Due to using Sobol sequences each subsequences must be \emph{uniform} and this protocol gives very good performance result to detect errors in the file blocks. In addition it detect error blocks very quickly. Transitition diagram~\ref{fig:tran1} shows the algorithm 1 and diagram~\ref{fig:tran2} describes about Algorithm 2. After partitioning the sequences, Coordinator will send the subsequences, $Sub_{i}$, to each $SUBTPA_{i}$.
 
Nevertheless, for sending $Sub_{i}$ to $SUBTPA_{i}$, from Coordinator takes extra network bandwidth. Although, it can not take any extra care about the critical data. To reduce the bandwidth usage and increase the efficiency by taking extra care about critical data, we device the Task Distribution Key ($\mathcal{TDK}$) to divides the sequences to subsequences. Our remaining protocols, describes about $\mathcal{TDK}$ based techniques in more details.

\section{$\mathcal{TDK}$ Based Distibution Scheme}
In $\mathcal{TDK}$ based scheme, we uses distinct Task Distribution Key($\mathcal{TDK}$) for each SUBTPA. $\mathcal{TDK}$s are binary strings consisting 0s and 1s. We uses $\mathcal{TDK}$ on the generated sequences as mask and takes the block numbers from sequences corresponding to 1 in the $\mathcal{TDK}$. 

Here, we uses two types of Task Distribution Key for interpret the the task at each SUBTPA side, one is \emph{ Non Overlapping $\mathcal{TDK}$} and another is \emph{Overlapping $\mathcal{TDK}$}. In case of Non-Overlapping $\mathcal{TDK}$s, the 1s are not overlap among different $\mathcal{TDK}$s. In our protocols we decides the $\mathcal{TDK}$ length from the number of SUBTPAs and number of 1s each $\mathcal{TDK}$s will contain. Here, we have shown one example, where four SUBTPA, and each contains four 1s. Hence, the $\mathcal{TDK}$ length should be calculated by multiplying number of SUBTPA and number of 1s for each $\mathcal{TDK}$s.
 \begin{equation*}
 \begin{split}
 1010000000001001\\
 0100011000100000\\
 0000100101000100\\
 0001000010010010
 \end{split}
\end{equation*} 
In case of Overlapping $\mathcal{TDK}$s, some 1s are overlap inside $\mathcal{TDK}$s. The Coordinator will decides how much amount of overlap needs among $\mathcal{TDK}$s. We repeat the above example wth 20\% overlap among $\mathcal{TDK}$s. Here, we can see that 4 bits are overlap among all four $\mathcal{TDK}$s.
 \begin{equation*}
 \begin{split}
 1010000010001001\\
 0100011000100001\\
 1000100101000100\\
 0001001010010010
 \end{split}
\end{equation*}  
When verify the file blocks those are not much critical, we can use Non-Overlapping $\mathcal{TDK}$s. But when we have to verify critical data we uses Overlapping $\mathcal{TDK}$s to increase the security of the stored data into Clouds. Thus, depends on the applications we have to choose which types of $\mathcal{TDK}$s we shall use. We gives the steps for generating non-overlapping TDK as follows:\\
\subsection{Non-overlapping Task Distribution Key generation}
Here, we formally describes the steps needed to construct the random Non-overlapping Task Distribution Key those are the heart of our $\mathcal{TDK}$ Based Distibution Scheme.
\begin{enumerate}
\item[1.] First Coordinator will decides the total number of SUBTPAs $n$ and the number of $1's$, $t$, which each $\mathcal{TDK}$ will contain.
\item[2.] After multiplying total number of SUBTPAs with $t$, Coordinator will generates the $\mathcal{TDK}$ length, $TDKLen$.
\item[3.] For placing the $1's$ inside each TDK, generates random permutation index from $1, 2,\ldots,TDKLen$.
\item[4.] Now takes first $t$ index from the permutation index set and place $1's$ for these indices into the first $\mathcal{TDK}$, after that takes next $t$ indices for second $\mathcal{TDK}$ and places $1's$ inside the second $\mathcal{TDK}$ corresponding to the $t$ indices. This process will continue up to all $\mathcal{TDK}$ will be generates.
\item[5.] After generating the $\mathcal{TDK}$s, if the length is co-prime to the sequence length, then Coordinator will distributes to SUBTPAs. Otherwise, Coordinator will adjust the $\mathcal{TDK}$ length to the next nearest primes or next Co-primes to the sequence length, $SeqLen$, to maintain the uniformity among subtasks.
\item[6.] The generated Task Distribution Keys ($\mathcal{TDK}$s) for $n$ SUBTPAs are $\sigma_{{S}_{1}}$, $\sigma_{{S}_{2}}$,  $\sigma_{{S}_{3}}$,$\ldots$, $\sigma_{{S}_{n}}$.
\end{enumerate}
This random $\mathcal{TDK}$ will generates at the Coordinator side and distributed among SUBTPA by using String Reconciliation.

\subsection{Overlapping Task Distribution Key generation}
To generate the Overlapping $\mathcal{TDK}$, first generates the Non-Overlapping $\mathcal{TDK}$ and need some extra adjustments. We describes the steps as follows:
\begin{enumerate}
\item[1.] After generating the Non-overlapping $\mathcal{TDK}$s, Coordinator decides the \% of overlap needs. 
\item[2.] Generates Random Permutation index of size same as \% of overlap within $1$,$2$,$\ldots,$ $TDKLen$ and place $1's$ according to the permutation index inside $\mathcal{TDK}$ where 0's present previously.
\end{enumerate}
If we want to verify critical data, we can use Overlapping $\mathcal{TDK}$, it provides strong integrity than Non-Overlapping $\mathcal{TDK}$ but takes little extra burden at the SUBTPA and Cloud Serverside.

\section{Approach for Protocol2}
Here, the Coordinator will randomly generates $\mathcal{TDK}$ and distributes among various SUBTPAs. Each SUBTPA will successively apply their $\mathcal{TDK}$ on the generated Sobol sequences as a mask up to the sequence will exhaust and take the corresponding sequence number as block number for verification.

As an concrete example, consider the $\mathcal{TDK}$ for the SUBTPA1 and SUBTPA2 are 10101 and 01010 respectively. Let, the generated Sobol random sequence is \{1216, 5312, 3264, 7360, 704, 4800, 2752, 6848, 1728\}, where file blocks are numbered from 0 to 8191. If we place the $\mathcal{TDK}$ for SUBTPA1 on the left end of the generated sequence and takes the block numbers corresponding to the 1, after that we slides the string to the right to the same length of the $\mathcal{TDK}$ and apply the same procedure then it generates the subtask for SUBTPA1 in (\ref{eq:sub1}) and similarly for SUBTPA2 in (\ref{eq:sub2}).\\
\{1216, 5312, 3264, 7360, 704, 4800, 2752, 6848, 1728\}\\
\hspace*{0.5cm}1\hspace{0.8cm}0\hspace{0.9cm}1\hspace{0.8cm}0\hspace{0.6cm} 1\\
 \{1216, 5312, 3264, 7360, 704, 4800, 2752, 6848,1728\}\\
  \hspace*{5.5 cm}1\hspace{0.7cm} 0\hspace{0.9cm}1\hspace{1cm}0\hspace{0.35cm}1 \\
\begin{equation}\label{eq:sub1}
\{ 1216, 3264, 704, 4800, 6848\}\\
\end{equation}
\{1216, 5312, 3264, 7360, 704, 4800, 2752, 6848, 1728\}\\
\hspace*{0.5cm}0\hspace{0.8cm}1\hspace{0.9cm}0\hspace{1cm}1\hspace{0.6cm} 0\\
\{1216, 5312, 3264, 7360, 704, 4800, 2752, 6848, 1728\}\\
\hspace*{5.6 cm}0\hspace{0.7cm} 1\hspace{0.8cm}0\hspace{0.8cm}1\hspace{0.4cm}0 \\
\begin{equation}\label{eq:sub2}
\{5312, 7360, 2752, 1728\}
\end{equation}

In our protocols, we uses two types of $\mathcal{TDK}$ for uniformly distributes the task among SUBTPAs and sometimes, we adjust the $\mathcal{TDK}$ length to balance the subtask for each SUBTPA.

\section{Protocol 2: Simple $\mathcal{TDK}$ Based Distribution Scheme}

\begin{figure}[!thbp]
 \centering
\subfigure[Before $\mathcal{TDK}$ Reconciliation]{
   \includegraphics[scale=1] {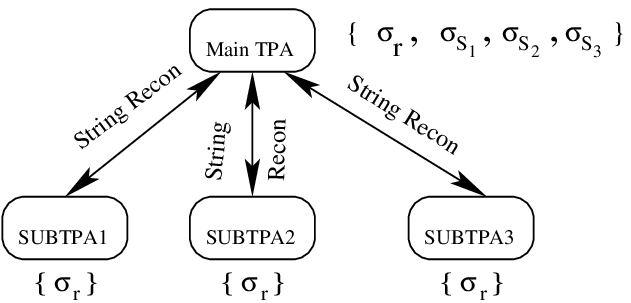}
   \label{fig:subfig1}
 }
 \subfigure[After $\mathcal{TDK}$ Reconciliation]{
   \includegraphics[scale=1] {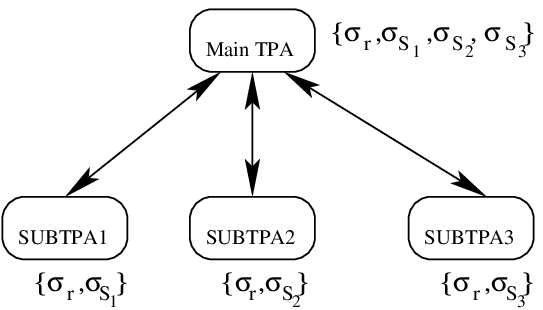}
   \label{fig:subfig2}
 }
\caption{$\mathcal{TDK}$ Exchange By Using String Reconciliation Protocol}
\label{fig: str}
\end{figure}

In our second protocol, Coordinator and each $SUBTPA_{i}$ will know the Sobol random key, $\sigma_{r}$, for generating the Sobol random sequences. In each new verification, Coordinator will  decides the parameters to generates the Sobol Random Key, $\sigma_{r}$ and publicly send to all $SUBTPA_{i}$. In addition, Coordinator generates $n$ number of random $\mathcal{TDK}$, $\sigma_{S_{i}}$, and distributes among $n$ SUBTPAs by using \emph{String Reconciliation Protocol} with some modifications [3].\\ 
Here, Coordinator knows $\sigma_{r},\sigma_{S_{1}}, \sigma_{S_{2}},\ldots,\sigma_{S_{n}}$ and each $SUBTPA_{i}$ know only $\sigma_{r}$, thus for reconciling, each $SUBTPA_{i}$ will perform the maximum computation(Characteristic Polynomial interpolation~\cite{setrecon}) needs for string reconciliation. The Coordinator maintains individual \emph{Modified deBruijn} graph for each $\sigma_{S_{i}}$, and for Sobol Random Key, $\sigma_{r}$, and each $SUBTPA_{i}$ maintains only Modified deBruijn graph for $\sigma_{r}$. We have shown, one diagram in figure~\ref{fig: str}, illustrates this communications.
After reconciling each $SUBTPA_{i}$ will know their $\mathcal{TDK}$, $\sigma_{S_{i}}$. Therefore, sending the $\sigma_{S_{i}}$ to $SUBTPA_{i}$, takes minimum communication due to String Reconciliation. Now, each $SUBTPA_{i}$ will generates Sobol Random Sequences and interpret their subsequence by using their own $\mathcal{TDK}$. We gave key and $\mathcal{TDK}$ generation and distribution of key and $\mathcal{TDK}$ in Algorithm 3. Algorithm 4, describes about the distributed challenge and verification for protocol 2. 

{\small
\begin{algorithm}[!htb]
\label{algo:Protocol2}
\SetCommentSty{sf}
 \caption{Key Generation and Distribution for Protocol2}

\Indp 
Coordinator randomly chooses one Primitive Polynomial, $\mathcal{P}$ of degree d and initialization number $m_{i}$, $i\in\{1,2,\ldots,d\}$ \; 
Coordinator decides Sobol Random Key, $\sigma_{r}=\langle\mathcal{P},m_{i},SKIP,LEAP,CONSTANT,SeqLen\rangle$\\
Coordinator Determines the Number of SUBTPAs, $n$, and threshold value, $m$ \;
Coordinator send $\sigma_{r}$ to all $SUBTPA_{i}$\;
Determine number of $1's$, $t$, each $TDK$ will contain\;
$TDKLen \leftarrow n \times t$\; 

Generates Random Permutation index from $1, 2, \ldots,TDKLen$\;
\For{$i \leftarrow 1$ to $n$}
{
    \For{$j \leftarrow 1$ to $TDKLen$}
     {
        $TDK[i][j] \leftarrow 0$\;
      }
}
 $i\leftarrow 1$\;
 $k \leftarrow 1$\;
\While{$i < n$}
{
    \For{ $j \leftarrow 1$ to $t$}
    {
       $l\leftarrow RandPerm[k++]$\;
       $TDK[i][l] \leftarrow 1$\;
    }
 $i \leftarrow i+1$\;
}
\eIf{ $gcd(TDKLen, SeqLen) \leftarrow 1$ }
{
   TDKLength is acceptable\;
} 
{
   TDK Length adjust to the next nearest Primes\;
}
Generated $TDK_{i}$ for $SUBTPA_{i}$ are represented as, $\sigma_{{S}_{1}}, \sigma_{{S}_{2}},  \sigma_{{S}_{3}},\ldots, \sigma_{{S}_{n}}$ respectively, distributes among SUBTPAs by using \emph{String Reconciliation Protocol}.\;

\Indm
\end{algorithm}

\begin{algorithm}[!htb] 
\label{proc:chal1}
\caption{DistributedChalandProofVerification for Protocol2}
\Indp
Each $SUBTPA_{i}$ generates  Sequence
\[\mathcal{L} \leftarrow f_{\sigma_{r}}(SeqLen)\]\\
Multiply CONSTANT \emph{powers of 2} with $\mathcal{L}$, to make each element as integer block number.\;
Interpret subsequence by using $\sigma_{S_{i}}$ as $r_{i,j}$
where $$\mathcal{L} \leftarrow \bigcup_{\substack{i\in[1,\ldots,n] \\ j\in[1,\ldots,p]}} r_{i,j}$$\\

$SUBTPA_{i}$ Calculates 10\% of $r_{i,j}$\;

\For {each $SUBTPA_{i}$}
{
   $l\leftarrow length(r_{i,j})$\;
   $Counter_{i} \leftarrow \lfloor (10/100)*l\rfloor$\;
} 

\For {$k \leftarrow 1$ to 10 }
{
   \For {$S\leftarrow1$ to $Counter_{i}$ and $t\leqslant l$}
   {
      $Chal_{i,k}[s] \leftarrow r_{i,j}[t]$\;
    }
    Send $\langle Chal_{i,k}\rangle$ to $SUBTPA_{i}$\;
    Wait for the proof, $PR_{i,k}$ from any Cloud Server\;
    $PR_{i,k} \leftarrow Receive()$\;
     \eIf{$PR_{i,k}$ equals to Stored Metadata}
      {
         $Report[k] \leftarrow TRUE$\;
      }
      {
         $Report[k] \leftarrow FALSE$\;
         Send Signal, $\langle Packet_{i,k}, FALSE\rangle$ to the Coordinator\;
       }
}

\Indm
\end{algorithm}
}

\subsection{Analysis of Protocol 2}
In $\mathcal{TDK}$ generation phase, we takes the mask length as co-prime to sequence length or prime length, because after applying $\mathcal{TDK}$ on $\mathcal{L}$, subsequences, $r_{i,j}$ becomes nonuniform, and to make it uniform, we uses these adjustments.
In algorithm 2 Coordinator generates Sobol Random Key and send to the SUBTPAs. In addition sends different $\mathcal{TDK}$, $\sigma_{S_{i}}$, for each $SUBTPA_{i}$. In this algorithm, we only uses Non-Overlapping $\mathcal{TDK}$ but we can use Overlapping $\mathcal{TDK}$ also. In DistributedChalandProofVerification2, $SUBTPA_{i}$ generates the Sobol random sequences by using key, $\sigma_{r}$ and stored in $\mathcal{L}$. Then, each $SUBTPA_{i}$ interpret their task by using corresponding $\mathcal{TDK}$, $\sigma_{S_{i}}$, and we denoted subtask for $SUBTPA_{i}$ as $r_{i,j}$ and defined as
\[\mathcal{L}=\bigcup_{\substack{i\in[1,\ldots,n] \\ j\in[1,\ldots,p]}} r_{i,j}\]
where 
 {\small \[p=\frac{SequenceLength}{TDKLength}+\xi\]}
   {\small $\xi=Number\, of\, 1's ~in \,first\,\\(SequenceLength~\%~TDKLength)\,length\, in\, TDK$}\;

Then, $SUBTPA_{i}$ will calculates 10\% of $r_{i,j}$ and creates challenge, $Chal_{i,k}$ and send to the server and waits for the proof, $PR_{i,k}$. After receiving the proof $SUBTPA_{i}$ will verify with the stored matadata and if the proof is correct then store TRUE in its table and if not match then store FALSE and send a signal immediately to the Coordinator for corrupt file blocks. Coordinator will receive \emph{signals} from any subset of $m$ out of $n$ SUBTPAs and ensures the fault location or stop the Audit operation. In the final step, Main TPA will gives the Audit result to the Client.

\section{Protocol 3:  $\mathcal{TDK}$ Based with artificial intelligence}
In this protocol, key generation and distribution phase is same as protocol 2. Here we change the Challenge and verification phase to enhance the performance of our protocol and at the same time reduce the computational overhead at the Coordinator side. We know that in Cloud Data storages, if file blocks will corrupt then there must be a chance to occur consecutively. Hence, when we generates randomly file block numbers then consecutive corrupt block numbers may appear in the first portion of the sequences and some in the last portition. In some situation one SUBTPA has sequences where corrupt block number appear in the first portion and anothers in the last portition. Thus, to detect consecutive errors takes too much time and send too many communications to the Cloud Server. Because, after detecting one error first time and sends a signal to Coordiantor, and another errors in the later and also sends signal to the Coordinator. Therefore, for detecting consecutive errors takes much more times and sends many signal to the coordinator. Instead of sending 10\% subsequence in a regular manner, if after detecting one block error, sends all the blocks those are not checked and near to that error blocks as a challenge provides fast error detection scheme. \\ 
To detect the consecutive errors very fast for each SUBTPA and reduce the communication cost to send a signal, in addition reduce Coordinator side computation, we change the verification and send signal portion of ChalandProofVerification for protocol 2 and gave it in algorithm 5 . This protocol uses $\mathcal{TDK}$ with Non-Overlapping and Overlapping mode. We describes the changed steps formally in the following:
\begin{enumerate}
 \item verify the received proof with the stored metadata. If match then send next 10\% subsequences.
\item If not matched then creates challenges with block numbers those are near to the corrupt block from the list, $r_{i,j}$ , and send to the server.
\item There may be a high probability to detect corrupt blocks insides this list. Finally SUBTPA will decides the range of error and send a signal to the Coordinator 
Here signal contains error block numbers and in which packets these are detected. Here packet means 10\% subsequence block numbers.
\item For choosing nearby blocks SUBTPA can take a predifined range mentioned by the Coordnator.
\end{enumerate}

{\small
\begin{algorithm}[!htb] 
\label{proc:chal}
\caption{DistributedChalandProofVerification for Protocol 3}
\Indp
Each $SUBTPA_{i}$ generates  Sequence
\[\mathcal{L} \leftarrow f_{\sigma_{r}}(SeqLen)\]\\
Multiply CONSTANT \emph{powers of 2} with $\mathcal{L}$, to make each element as integer block number.\;
Interpret subsequence by using $\sigma_{S_{i}}$ as $r_{i,j}$
where $$\mathcal{L} \leftarrow \bigcup_{\substack{i\in[1,\ldots,n] \\ j\in[1,\ldots,p]}} r_{i,j}$$\\

$SUBTPA_{i}$ Calculates 10\% of $r_{i,j}$\;

\For {each $SUBTPA_{i}$}
{
   $l\leftarrow length(r_{i,j})$\;
   $Counter_{i} \leftarrow \lfloor (10/100)*l\rfloor$\;
} 

\For {$k \leftarrow 1$ to 10 }
{
   \For {$S\leftarrow1$ to $Counter_{i}$ and $t\leqslant l$}
   {
      $Chal_{i,k}[s] \leftarrow r_{i,j}[t]$\;
    }
    Send $\langle Chal_{i,k}\rangle$ to $SUBTPA_{i}$\;
    Wait for the proof, $PR_{i,k}$ from any Cloud Server\;
    $PR_{i,k} \leftarrow Receive()$\;
     \eIf{$PR_{i,k}$ equals to Stored Metadata}
      {
         $Report[k] \leftarrow TRUE$\;
      }
      {
         $Report[k] \leftarrow FALSE$\;
         Send Signal, $\langle Packet_{i,k}, FALSE\rangle$ to the Coordinator\;
       }
}

\Indm
\end{algorithm}
}

\subsection{Analysis of Protocol 3}
This protocol will detect errors very quickly rather that Protocol 1 and Protocol 2 and also this protocol reduces the communication cost between Coordinator and SUBTPA as well as SUBTPA and Cloud server side. But in this protocol one major problem is that SUBTPAs are centrally controlled by the Coordinator. Now in our next protocol Coordinator only generates the Sobol Random Key, $\sigma_{r}$, and send to all SUBTPAs. After that all challenge and verification steps are performed by the SUBTPA. We give this protocol in the next section.

\section{Protocol 4: $\mathcal{TDK}$ Based with Full Autonomy}
All the previous Protocols are depending on the Coordinator for interpret the generated sequence and get the subsequence for integrity verification. Hence, all are centrally controlled by the Coordinator. As, Cloud Computing is based on  \emph{Distributed Data and Distributed Data} paradigm, we are trying to model our last protol suitable for that model. Here, each SUBTPA will independently generates sequences and interprets any 10\% block number and send as a challenge to the Server. In the above protocols, Coordinator decides the subtask for each SUBTPA by using simple partition or  $\mathcal{TDK}$ based approach, but all these takes much time. Here each SUBTPA will independtly decides the $\mathcal{TDK}$ and interpret the sequence to get the subsequence. Therefore, SUBTPA has full autonomy about the varification scheme. So, when Coordinator wants to sense the error insides the file blocks very quickly the following protocols works very well.

\textbf{Key Generation and Distribution::}
\begin{enumerate}
\item[1.] Coordinator chooses one Sobol Random Key, $\sigma_{r}$ and publicly send to the SUBTPAs.\\
\textbf{Challenge generation and Proof Verification Phase::}
\item[2.] $SUBTPA_{i}$ will independently generates the Sobol random sequences by using same Sobol Random Function in~(\ref{eq: sobf}):
\begin{equation}\label{eq: sobf}
 \mathcal{L}=f_{\sigma_{r}}(i) 
\end{equation}
\item[3.]Each SUBTPA independently generates $\mathcal{TDK}$ for task generation.
\item[4.] Each SUBTPA independently interpret any 10\% of the file block number from the generated sequences and make a challenge and send to the Server.
\item[5.] Any Server sends proof to the requested SUBTPA.
\item[6.] Each SUBTPA will check the proof with the stored metadata. If it cannot match with metadata then SUBTPA will immediately sends a \emph{signal} to the Coordinator.\\
\item[5.] Coordinator will receive signals from any subset of $m$ out of $n$ SUBTPAs and ensures the fault location.
\end{enumerate}

\subsection{Analysis of Protocol 4}
In the above protocol SUBTPA has liberty to check any 10\% block from the generated sequences. Thus, there may be a chance to overlap among the SUBTPAs, but this protocol is very much helpful to detect the data corruption very first and also for critical data block, it will works better. For the above protocol, each time SUBTPA will get a mismatch in the proof verification it sends a signal to the Coordinator. Therefore this protocol is maintain the \emph{Distributed Control and Distributed Data} paradigm. 
 
Here, we generalizes the integrity verification protocol in a distributed manner. Therefore, we can use our protocols on existing RSA based~\cite{shyamrsa}~\cite{p16} or ECC~\cite{shyamecc} based protocol to make distributed RSA or ECC protocols. In the next section, we discusses about the implementation and analysis of the outcomes from various protocols.

\chapter{Implementation and Analysis of Results}
\label{ch:results}

In this chapter, we discusses about the implementation of our distributed verification protocols and various analysis on the outcome results. Finally, we device a mathematical model based on the analysis of the results.

\section{Implementation}
Now, we present and discusses about the implementation of our various protocols. To verify the performance of our protocols, we uses desktop with Core2 Duo 2.2 GHz processor, 2GB RAM, and 500 GB SATA Hard Drive. All programs are written in C language on Linux OS. We used MATLAB R2009a software for generating Sobol Random Sequences~\cite{sobol1} and for analysis of the result.

In our protocols, we uses Sobol random sequence generator to generate the file block numbers, because generated  sequences are uniformly distributed over [0, 1) and cover the whole range. This property will holds for any length powers of two. But, if we take the sample sequence with any length, then also holds uniformity with little fluctuations and that does not create any problem for our proposed protocols. We saw that, in case of, Pseudo Random number generator, the generated sample sequence does not satisfy the uniformity property. Due to uniformness of the Sobol sequence it is very easy to detect nondeterministic(unknown) error quickly rather than Pseudo Random sequence.

For generating the Sobol sequence we need one Sobol random key and depending on the key the sequence will change. Hence, by using different key we can generates different sequence for each verification in our protocols. We mention the parameters for constructing the key in chapter 3 and the sequence generation procedure in Chapter 2, here, we restate about key for completeness of our discussion. 

\[\sigma_{r}=\langle\mathcal{P},m_{i},SKIP,LEAP,CONSTANT,SeqLen\rangle\]

Where, $\sigma_{r}$ is the Sobol random key that contains one primitive polynomial, $\mathcal{P}$ of degree $d$. The number of Primitive polynomials are determined by using $\phi(2^{d}-1)/d$, where d is the degree of the polynomials. In Table~\ref{tb:pol} we gave a list of primitive polynomials upto degree 6. We discussed about the $m_{i}$ values in section 2.1 of Chapter 2. The SKIP and LEAP values are used to skip and leap on the generated sequence. The CONSTANT is very important because, the generated sequence values are fractions and by multiplying CONSTANT we convert to the integer block number values. The CONSTANT values should be powers of 2 and obviously it is same to the total number of blocks stored into the Cloud servers. The SeqLen defines the sample(sequence) size that will generates for each integrity verification.

\begin{table}[!htbp]

\begin{tabular}{|c|l|}
\hline
\textbf{Degree}&\textbf{Primitive Polynomials}\\
\hline
1& $1+x$\\
\hline
2&$1+x+x^{2}$\\
\hline
3& $1+x+x^{3}$, $1+x^{2}+x^{3}$\\
\hline
4& $1+x+x^{4}$, $1+x^{3}+x^{4}$ \\
\hline
\multirow{2}{*}{5}& $1+x^{2}+x^{5}$, $1+x+x^{2}+x^{3}+x^{5}$, $1+x^{3}+x^{5}$,\\
 &$1+x+x^{3}+x^{4}+x^{5}$, $1+x^{2}+x^{3}+x^{4}+x^{5}$, $1+x+x^{2}+x^{4}+x^{5}$\\
\hline
\multirow{2}{*}{6}& $1+x+x^{6}$, $1+x+x^{2}+x^{5}+x^{6}$, $1+x^{2}+x^{3}+x^{5}+x^{6}$,\\
&$1+x+x^{3}+x^{4}+x^{6}$, $1+x+x^{4}+x^{5}+x^{6}$, $1+x^{5}+x^{6}$\\
\hline
\end{tabular}
\caption{List of Primitive Polynomials}
\label{tb:pol}
\end{table}

In addition, in our Distributed verification protocol 1 Main TPA/Coordinator will decides a random key and generates the Sobol random sequence for verification and distributes among SUBTPAs based on simple partition. We can say from the property of Sobol sequence that each subsequence also maintain uniformity. Therefore, not only one SUBTPA will detect errors but also all the SUBTPAs concurrently detect errors and increase the performance of the verification scheme. Since, each subtask should be balance for each SUBTPA. Let, the number of file blocks is 10000, and logically partition into four consecutive segments, and if Coordinator generates a sample sequence of length 128 and distributes among 4 SUBTPAs, then each has 32 block number. Hence, each SUBTPA must ensures that it will verify 8 blocks from each segment out of four, but in case of Pseudo Random sequence we cannot ensures this fact. If the sample size is 130 then 3 SUBTPA has 32 block numbers and fourth will take 34 block number for verification and does not create any problem for uniformness.
 
\begin{figure}[tbh]
\centering
\includegraphics[height=3in,width=4.5in]{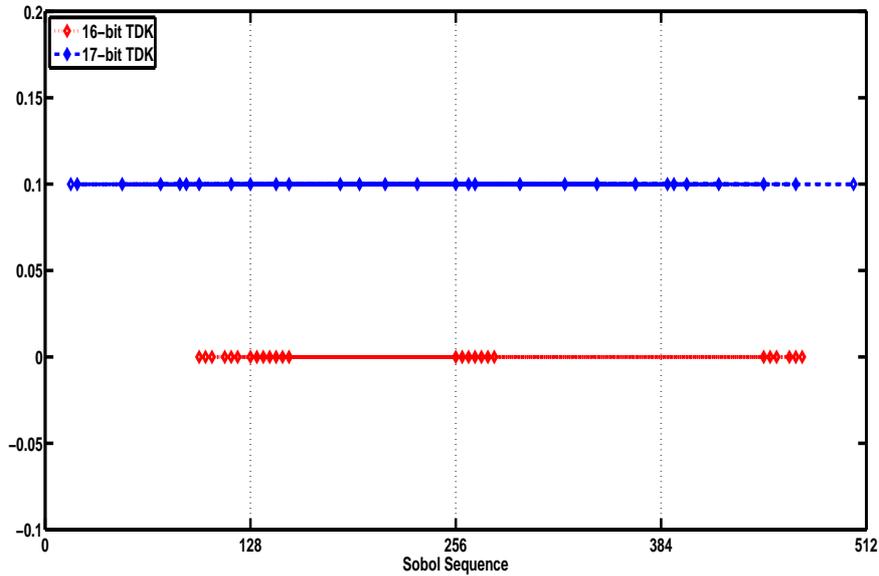}
\caption{16-bit $\mathcal{TDK}$ Extends to 17-bit}
\label{fig: uniform}
\end{figure}

For remaining protocols, Main TPA will decides the common Sobol Random key and a set of random $\mathcal{TDK}$ for each SUBTPA and distributes among SUBTPAs. Then each SUBTPA will generates the random sequence and apply the $\mathcal{TDK}$ given by Main TPA to interpret the subsequence from the generated sequence. In $\mathcal{TDK}$ approach after interpreting the generated subsequence may be nonuniform because we uses random binary string consisting $0's$ and $1's$ as a $\mathcal{TDK}$.
We observed that, if $\mathcal{TDK}$ length is Co-Prime to the sequence(samples) length then the generated subsequence maintain uniformity for each SUBTPA. But, if the $\mathcal{TDK}$ length is powers of 2 then the subsequences are forms cluster for each SUBTPA and as a result \emph{nonuniform} subtask, to handle this problem we adjust the $\mathcal{TDK}$ length to the nearest \emph{prime or Co-prime to the sequence length}. So that, after applying powers of 2 length $\mathcal{TDK}$ the subsequence must maintain uniformness. Figure~\ref{fig: uniform} illustrates our observation for only one subtask, if any SUBTPA will takes 16-bit $\mathcal{TDK}$ for any Sobol sequences, then the sutask becomes nonuniform but if it extends to 17 bits and apply on the same sequences then the generated subtask should maintain uniformity.

Therefore, finally we can say that all our protocols holds the uniformity property for integrity verification.

In addition, we saw that in case of Sobol Random sequence number of error detected by the protocols are constant. That means, by using same sample length with different sequence the number of error detected is approximately fixed. In the next section, we describes the outcomes from different protocols and analysis the results.  

\section{Output Analysis}
It is very natural that audit activities would increase the communication and computational overheads of audit services. To enhance the performance, we used the String Reconciliation Protocol to distribute the $\mathcal{TDK}$ that provides minimum communication and tractable computational complexity. Thus, reduces the communication overhead between Main TPA and SUBTPAs. For each new verification, Coordinator can change the $\mathcal{TDK}$ for any SUBTPA and send only the difference part of the multiset element to the SUBTPA. In addition, we used probabilistic verification scheme based on Sobol Sequences that provides not only uniformity for whole sequences but also for each subsequences, so each SUBTPA will independently verify over the whole file blocks. Thus, there is a high probability to detect fault location very quickly. Therefore, Sobol sequences provides strong integrity proof for the remotely stored data. 

\begin{figure}[tbh]
\centering
\includegraphics[height=3in,width=4.5in]{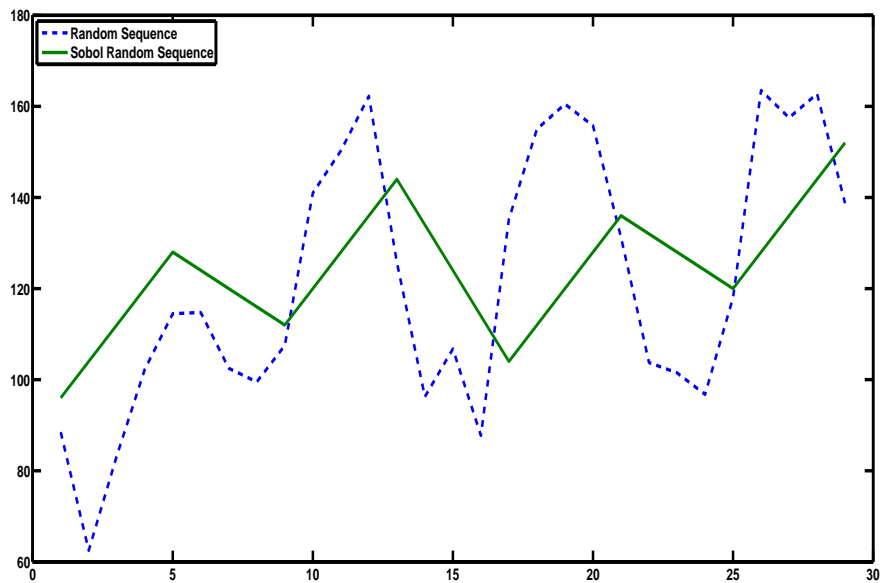}
\caption{Comparison  between Successive Mean of Random and Sobol Random Sequences}
\label{fig:amean}
\end{figure}

Here, we consider one concrete example, taking 32 numbers from the Sobol sequences and general Pseudo Random sequences and takes consecutive four numbers successively, calculates the arithmetic mean and have shown in figure~\ref{fig:amean}.

We can see from figure~\ref{fig:amean}, that the arithmetic mean of the consecutive Sobol sequences span from 100-150 but in case of random number it span from 60-160. It imply that, in case of Sobol Sequences the uniformity property will hold for any consecutive sequence numbers.

Now, we are giving the outputs of our protocols. For first setup, we took 10,48,576 file blocks, and 1\% error blocks(10,485). We generates 20\% sample size(209715) out of 10,48,576 file blocks. Error block numbers are generated randomly. We takes 20 SUBTPAs and distributes the sample sequence among them. We took 25 trials with different sample sequence and different random error blocks, and gives the result in the Table 4.2, where each column represents one trial and each row for one SUBTPA, intersection of row and column element defines total number of errors detected by one SUBTPA for one trial. First row of this table defines the trial number and first column represents the SUBTPA number. Very interesting is the last row, that defines the total number of errors detected by all SUBTPAs for a single trial. More deeply, if we observe the last row, we can see that total number of errors detected by all SUBTPAs for a single trial is 20\%(2100) of the total error blocks(10485). We saw that, this fact is true for any case. If total number of errors are 0.5\% then 20\% sample size detect upto 40\% error blocks and if error is 1\% and sample size is 30\% of the total blocks then detect upto 30\% error blocks out of 10,485 error blocks. In addition we have observed that all the SUBTPAs are detected the First error in first packet(first 10\% subsequence).

Now we can construct another table in~\ref{tb:firsttest2} from the above where, each row for one SUBTPA and Max. columns gives the maximum number of error blocks detected out of 25 trials and Min. columns gives the minimum number of error blocks detected out of 25 trials for a corresponding SUBTPA. Fourth columns Avg. gives the average number of errors detected out of 25 trials and Standard Deviation columns gives the standard deviation of the detected error. From this table we can see the minimum number of error detected by each SUBTPA is approximately 85 and maximum error detected approximately 124. Therefore we can set the threshold value for each SUBTPA so that when Coordinator will detect m number of errors out of n number of errors from one SUBTPA then Coordinator will stop the verification process.
\begin{table*}[tbh]
\begin{center}
\begin{tabular}{|c|c|c|c|c|}
\hline
\textbf{SUBTPA}&\textbf{Max.}&\textbf{Min.}&\textbf{Avg}&\textbf{Standard Deviation}\\
\hline
S1&117&88&100&8.8365\\
\hline
S2 &119&86&104&8.3652\\
\hline
S3& 128&89&105&12.0634\\
\hline
S4& 124&80&104&10.9435\\
\hline
S5&113&91&102&7.6974\\
\hline
S6&123&83&104&9.7386\\
\hline
S7&120&80&100&11.5794\\
\hline
S8&121&90&107&7.8926\\
\hline
S9&128&88&104&10.0876\\
\hline
S10&132&91&105&10.5906\\
\hline
S11&121&96&105&6.8981\\
\hline
S12&131&82&108&11.4110\\
\hline
S13&125&91&107&10.2823\\
\hline
S14&132&85&104&13.3944\\
\hline
S15&137&84&105&14.1372\\
\hline
S16&129&80&109&11.8286\\
\hline
S17&128&82&103&10.8.39\\
\hline
S18&126&89&106&10.5920\\
\hline
S19&129&86&104&12.3237\\
\hline
S20&125&82&104&10.9716\\
\hline
\end{tabular}
\end{center}
\caption{Protocol 1:Report Table with Modified form}
\label{tb:firsttest2}
\end{table*}
Now we slidely modify the Table~\ref{tb:firsttest2} and write in another form and has given in Table~\ref{tb:firsttest3}. In this table $S_{i}$ represents $SUBTPA_{i}$.

\begin{table*}[tbh]
\begin{center}

\begin{tabular}{|c|c|}
\hline
\textbf{Min. Range}&\textbf{SUBTPAs}\\
\hline
80-85&$S_{4}$,$S_{6}$,$S_{7}$,$S_{12}$,$S_{14}$,$S_{15}$,$S_{16}$,$S_{17}$,$S_{20}$\\
\hline
86-90&$S_{1}$, $S_{2}$,$S_{3}$,$S_{8}$,$S_{9}$,$S_{18}$,$S_{19}$\\
\hline
91-96& $S_{5}$,$S_{10}$,$S_{11}$,$S_{13}$\\
\hline
\end{tabular}
\end{center}
\caption{Protocol 1:Report Table with SUBTPA Class based on Minium Error Detection}
\label{tb:firsttest3}
\end{table*}

Now we are giving two tables, first one for protocol 1 and second is for protocol 2, where total number of SUBTPAs are 30 and total number of blocks is 41,94,304, sequence size is 20\% of total number of blocks, total number of error blocks is 1\% of 41,94,304. For protocol 2 we takes $\mathcal{TDK}$ size to 90, where each $\mathcal{TDK}$ contain 3 $1's$. We took 12 trials and for each trial we change the sequence and for some trial change the error block numbers. The result for protocol 1 has given in Table~\ref{tb:prot1} and for Protocol 2 in Table~\ref{tb:prot2}. From both table we can see that total number of error detected in each trial is 20\% of the 1\% error blocks. These two tables also shows the maximum, minimum, average, and standard deviation of the error detected by each SUBTPA. 

\begin{table}[tbh]
{ \tiny
\begin{tabular}{|c|c|c|c|c|c|c|c|c|c|c|c|c|c|c|c|}
\hline
\textbf{SUBTPA}&\textbf{T1}&\textbf{T2}&\textbf{T3}&\textbf{T4}&\textbf{T5}&\textbf{T6}&
\textbf{T7}&\textbf{T8}&\textbf{T9}&\textbf{T10}&\textbf{T11}&\textbf{T12}&
\textbf{Average}&\textbf{Max}&\textbf{Min} \\
\hline
S1&286&286&275&289&292&270&252&280&251&287&269&273&275.83&292&251\\
\hline
S2&314&314&290&274&286&266&289&256&274&272&304&271&284.17&314& 256 \\
\hline
S3&269&269&268& 273&272&299&293&276&292&307&287&290&282.92&307&268\\
\hline
S4&279&279&272&274&275&273&280&278&279&281&283&288&278.42&288&272\\
\hline
S5&279&279&276&292&284&275&293&283&260&299&270&311&283.42&311&260\\
\hline
S6&292&292&256&278&273&281&308&284&287&299&277&282&284.08&308&256\\
\hline
S7&266&	266&279&253&282&300&260&293&308&325&277&281&282.5&325&253\\
\hline
S8&288&	288&287&267&283&308&289&288&263&262&260&286&280.75&308&260\\
\hline
S9&251&251&271&306&309&278&281&256&295&241&279&274&274.33&309&241\\
\hline
S10&280&280&291&289&266&290&293&279&272&297&271&272&281.67&297&266\\
\hline
S11&284	&284&286&281&288&279&278&269&287&245&279&299&279.92&299&245\\
\hline
S12&286&286&257&273&291&280&283&276&271&304&283&255&278.75&304&255\\
\hline
S13&262&262&295&277&283&267&302&257&292&271&261&279&275.67&302&257\\
\hline
S14&284&284&297&277&284&296&273&270&274&276&296&261&281&297&261\\
\hline
S15&305&305&287&259&313&259&286&288&259&281&285&298&285.42&313&259\\
\hline
S16&270&270&298&278&265&269&283&296&259&280&256&281&275.42&298&256\\
\hline
S17&290&290&302&271&295&272&244&277&278&244&306&294&280.25&306&244\\
\hline
S18&270	&270&291&279&282&293&312&315&304&289&310&292&292.25&315&270\\
\hline
S19&288&288&252&284&308&275&320&270&289&277&285&285&285.08&320&252\\
\hline
S20&268&268&282&259&304&302&279&257&281&261&279&295&277.92&304&257\\
\hline
S21&299&299&275&295&276&287&256&271&287&282&273&281&281.75&299&256\\
\hline
S22&267&267&289&285&287&282&281&281&297&292&291&276&282.92&297&267\\
\hline
S23&265&265&273&257&279&301&266&263&284&288&312&298&279.25&312&257\\
\hline
S24&253&253&273&305&268&287&292&282&282&238&306&277&276.33&306&238\\
\hline
S25&284&284&277&308&282&268&255&302&285&314&287&279&285.42&314&255\\
\hline
S26&301&301&282&282&282&308&282&277&297&276&290&274&287.67&308&274\\
\hline
S27&283	&283&251&283&269&273&297&260&284&307&270&273&277.75&307&251\\
\hline
S28&284&284&291&274&273&275&282&304&293&306&311&267&287&311&267\\
\hline
S29&291&291&288&293&300&284&230&255&247&273&267&276&274.58&300&230\\
\hline
S30&305&305&251&310&270&267&287&298&292&284&275&270&284.5&310&251\\
\hline
Total&8443&8443&8362&8425&8521&8464&8426&8341&8423&8458&8499&8438&8436.92&9181&7685\\
\hline
\end{tabular}}
\caption{Protocol 1:Report Table for 12 Trials with 30 SUBTPAs}
\label{tb:prot1}
\end{table}

\begin{table}[tbh]
\centering
{\tiny
\begin{tabular}{|c|c|c|c|c|c|c|c|c|c|c|c|c|c|c|c|}
\hline
\textbf{SUBTPA}&\textbf{T1}&\textbf{T2}&\textbf{T3}&\textbf{T4}&\textbf{T5}&\textbf{T6}&
\textbf{T7}&\textbf{T8}&\textbf{T9}&\textbf{T10}&\textbf{T11}&\textbf{T12}&
\textbf{Average}&\textbf{Max}&\textbf{Min}  \\
\hline
S1&288&272&282&263&296&294&285&294&386&	278&284&270&291&386&263\\
\hline
S2&325&306&245&259&296&251&247&292&287&329&282&231&279.17&329&231\\
\hline
S3&273&261&308&290&293&268&302&259&281&302&342&266&287.08&342&259\\
\hline
S4&283&292&279&287&267&287&286&259&279&274&268&371&286&371&259\\
\hline
S5&312&282&289&264&299&292&294&285&268&296&292&285&288.17&312&264\\
\hline
S6&315&292&287&315&270&291&258&279&278&268&269&257&281.58&315&257\\
\hline
S7&293&271&285&304&268&297&267&267&286&244&248&304&277.83&304&244\\
\hline
S8&277&282&290&290&292&298&295&306&277&281&262&280&285.83&306&262\\
\hline
S9&273&274&263&268&266&259&285&258&272&300&316&282&276.33&316&258\\
\hline
S10&267&271&279&282&309&292&274&290&311&288&271&258&282.67&311&258\\
\hline
S11&260&288&271&280&260&291&281&299&261&287&270&265&276.08&299&260\\
\hline
S12&285&278&310&297&310&273&270&288&253&284&290&290&285.67&310&253\\
\hline
S13&287&278&268&238&262&280&278&286&284&274&306&258&274.92&306&238\\
\hline
S14&238&301&276&266&264&310&293&259&270&283&301&271&277.67&310&238\\
\hline
S15&295&243&272&276&287&279&287&270&288&276&264&301&278.17&301&243\\
\hline
S16&249&283&274&279&288&264&256&266&268&272&281&283&271.92&288&249\\
\hline
S17&291&285&291&277&288&266&271&281&286&308&264&302&284.17&308&	264\\
\hline
S18&255&291&281&277&273&310&277&294&298&248&280&280&280.33&310&248\\
\hline
S19&273&290&281&316&294&261&270&282&304&242&299&271&281.92&316&242\\
\hline
S20&278&292&255&283&289&261&299&264&294&281&293&267&279.67&299&255\\
\hline
S21&270&308&299&252&280&285&295&283&280&273&272&281&281.5&308&252\\
\hline
S22&279&285&261&277&299&251&285&267&276&256&293&308&278.08&308&251\\
\hline
S23&282&267&272&282&276&295&292&263&264&282&265&279&276.58&295&263\\
\hline
S24&279&292&285&279&275&272&305&292&294&297&300&272&286.83&305&272\\
\hline
S25&275&252&264&285&285&288&280&290&260&282&274&280&276.25&290&	252\\
\hline
S26&264&283&280&285&285&261&256&265&246&279&262&269&269.58&285&	246\\
\hline
S27&320&285&267&279&264&279&275&249&286&304&278&276&280.17&320&249\\
\hline
S28&266&288&280&299&263&316&269&282&278&282&283&296&283.5&316&263\\
\hline
S29&307&288&284&295&299&269&294&302&259&289&292&291&289.08&307&259\\
\hline
S30&284&263&284&281&307&324&300&288&277&272&283&294&288.08&324&263\\
\hline
Total&8443&8443&8362&8425&8504&8464&8426&8359&8451&8431&8484&8438&8435.83&9397&7615\\
\hline
\end{tabular}}
\caption{Protocol2: Report Table for 12 Trials with 30 SUBTPAs}
\label{tb:prot2}
\end{table}

\begin{figure}[tbh]
\centering
\includegraphics[height=3in,width=4.5in]{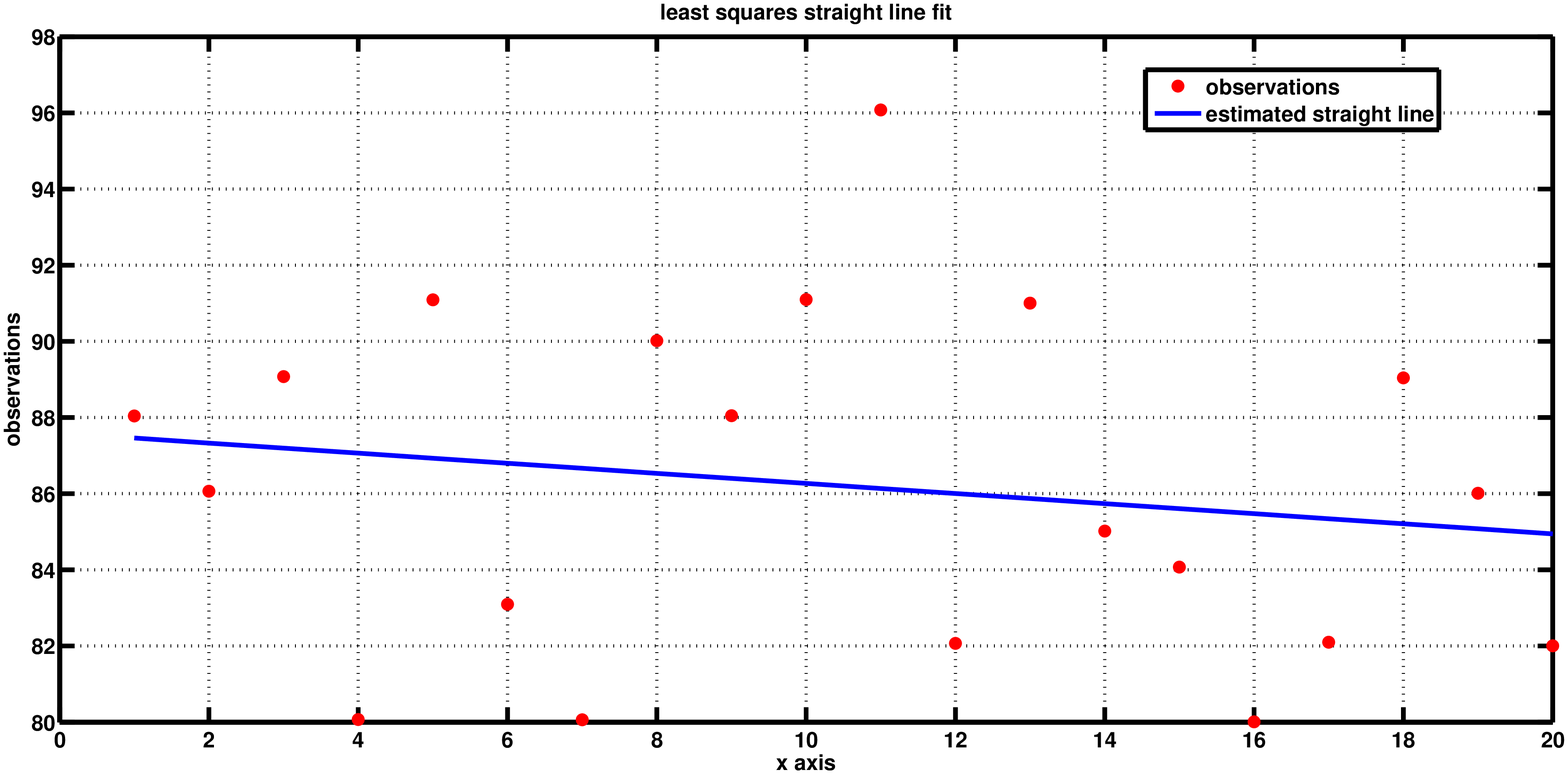}
\caption{Least Square line for 20 SUBTPA depends on minimum error detection}
\label{fig: lst1}
\end{figure}

\begin{figure}[bthp]
\centering
\includegraphics[height=3in,width=4.5in]{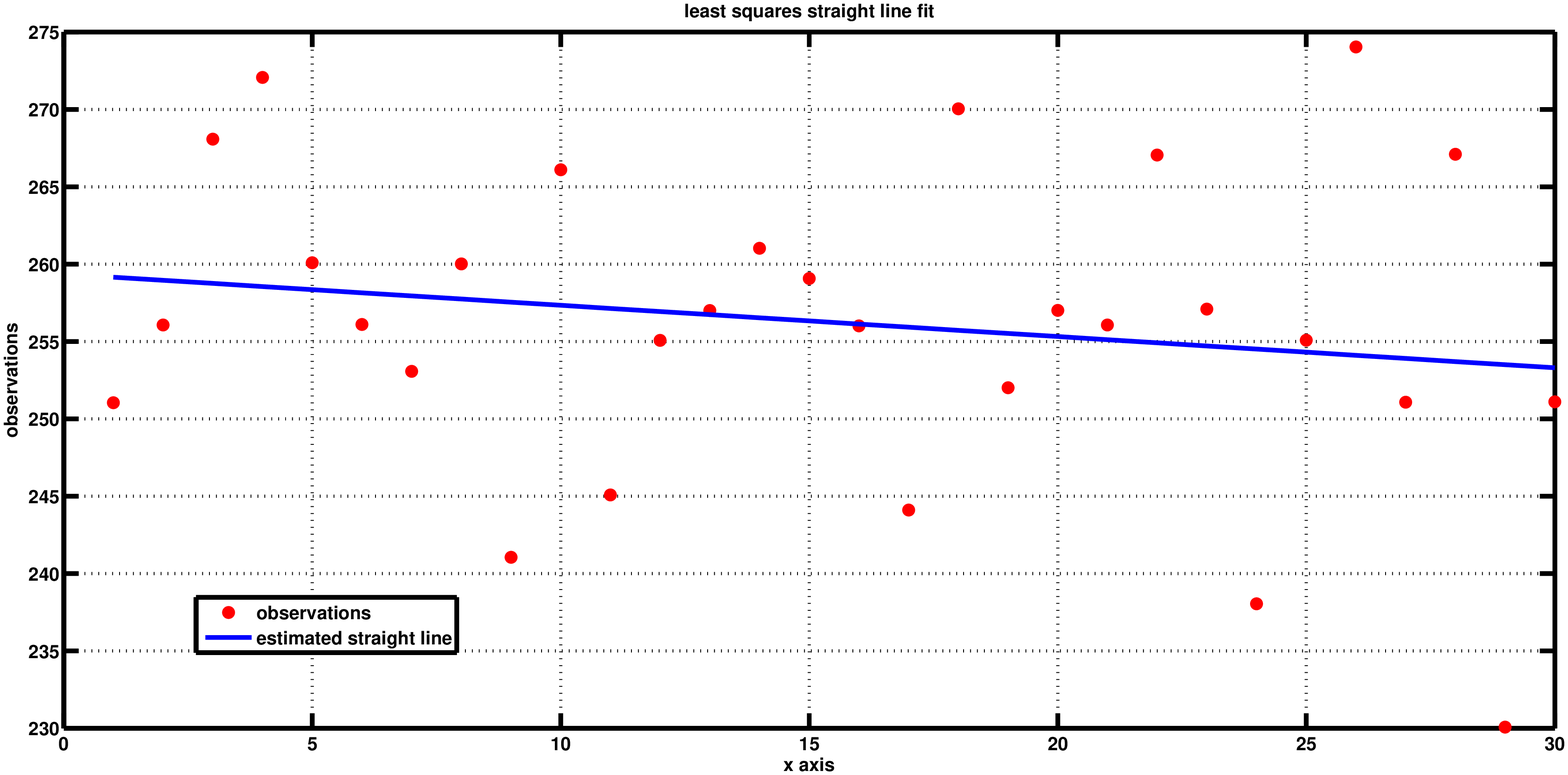}
\caption{Protocol1: Least Square line for 30 SUBTPA depends on average error detection}
\label{fig: lst2}
\end{figure}

\begin{figure}[!tbhp]
\centering
\includegraphics[height=3in,width=4.5in]{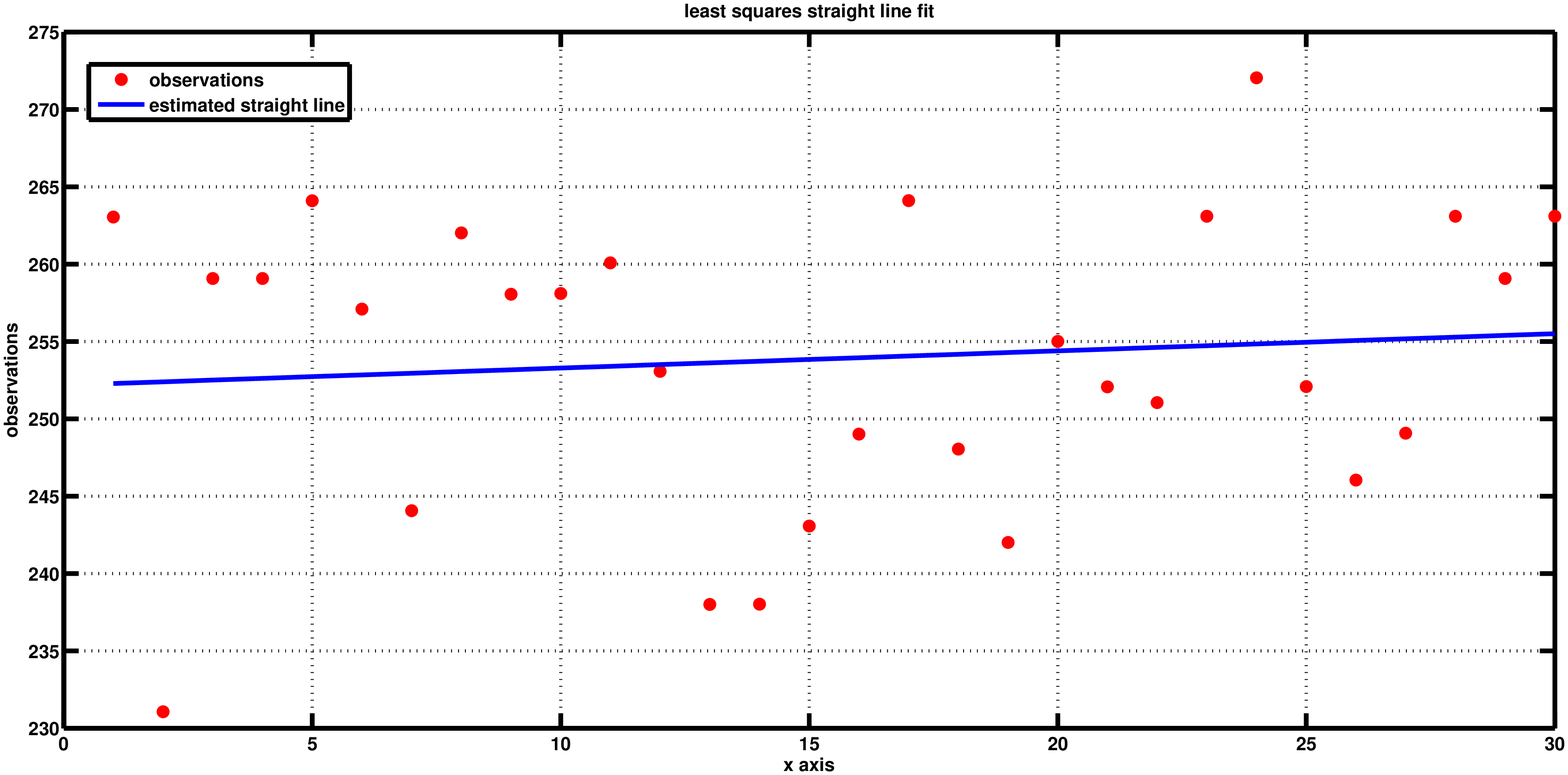}
\caption{Protocol2: Least Square line for 30 subtpa depends on average error detection}
\label{fig: lst3}
\end{figure}

Depending on the outcome from different trials we made above tables, shows the minimum, maximum and average number of errors detected by each SUBTPA. Now, we can fit the \emph{least square lines} corresponding to minium and average number of errors detected by different SUBTPAs. This shows the best fit line, a average range of error detected by each SUBTPAs. Here, we have shown least square lines for minimum and average error detected by protocol1 and average number of error detected by protocol2 in Figure~\ref{fig: lst1},~\ref{fig: lst2}, and ~\ref{fig: lst3} respectively. 


\newtheorem{thm}{Theorem}
\begin{thm}
Subtask for each SUBTPAs must be uniform after applying Task Distribution Key, if $\mathcal{TDK}$ length is prime or $\mathcal{TDK}$ length is relatively prime to the Sequence length.
\end{thm}
We know that Sobol Sequences are Quasi Random Sequences of the uniformity property. Now, if we generates random block numbers by using Sobol Random generator for a given length, then it must be uniform. In addition, if we simply partition the sequences into subsequences and distributes among various SUBTPAs, then each subsequence must be maintain the uniformity.
But when we uses Task distribution Key then subtask may or may not be uniform.
We saw that when the $\mathcal{TDK}$ length is powers of 2, then generated subtask does not maintain the uniformity property. Because, in Sobol sequence maintain some pattern, if we take 4 consecutive number then we can see that these numbers are from four regions over the Sequences, if we divide the full sequences into four regions, and for 8, 16, 32,\ldots it also hold. When we placed the $\mathcal{TDK}$ over the generated Sequences then Subtask contain those numbers whose corresponding $\mathcal{TDK}$ bit is 1 and successively applying this $\mathcal{TDK}$ to generates the sequences. Thus if the $\mathcal{TDK}$ length is powers of two then for each successive $\mathcal{TDK}$ shifting, the chosen block numbers must be very close to each other and forms cluster.
If we take $\mathcal{TDK}$ length as prime then in each successive shifting the chosen block numbers are spread over the segment. Therefore, maintains the uniformity for each subtask. Now if the $\mathcal{TDK}$ length is Co-prime means
\[gcd(TDKLength, SequenceLength)=1\]  
Then there is no factor equals to the powers of 2, that means for each successive $\mathcal{TDK}$ shifting block numbers are spread over the whole sequences and maintain the uniformity property for each subtask. Therefore, generated subtask must be uniform if the $\mathcal{TDK}$ length relatively prime to the sequence length and also hold for prime length.

\begin{table}[thbp]
{\footnotesize
\begin{tabular}{|c|l|}
\hline
\textbf{Mean Value}&\textbf{SUBTPAs}\\
\hline
\multirow{2}{*}{38-39}&S23,S120,S177,S8,S13,S15,S18,S32,S33,\\
&S37,S42,S51,S61,S90,S127,S129,S176,S196\\
\hline
\multirow{2}{*}{40}&S6,S26,S28,S36,S45,S54,S58,S59,S74,S75,S93,\\
&S102,S110,S125,S130,S135,S136,S143,S185,S190,S192,S199\\
\hline
\multirow{3}{*}{41}&S9,S12,S14,S19,S22,S35,S38,S49,S50,S53,S64,S65,S70,\\
&S77,S84,S97,S104,S105,S114,S115,S117,S118,S126,S128,\\
&S138,S139,S151,S154,S158,S161,S166,S170,S171,S174,S191,\\
\hline
\multirow{4}{*}{42}&S4,S24,S30,S40,S43,S44,S48,S56,S57,S62,S76,\\
&S80,S83,S91,S92,S95,S98,S103,S111,S112,S116,S122,\\
&S131,S132,S134,S137,S146,S153,S156,S164,S165,\\
&S168,S169,S173,S175,S179,S186,S187,S189,S193,S197,S200\\
\hline
\multirow{4}{*}{43}&S7,S11,S16,S20,S25,S27,S29,S31,S39,S41,\\
&S46,S47,S66,S67,S69,S71,S81,S82,S85,S96,S107,\\
&S113,S121,S124,S144,S147,S148,S149,S150,S155,\\
&S157,S163,S167,S172,S180,S183,S184,S194,S195\\
\hline
\multirow{2}{*}{44}&S1,S3,S5,S17,S60,S72,S73,S78,S79,S87,S94,\\
&S106,S109,S119,S142,S152,S159,S181,S182,S188,S198\\
\hline
\multirow{2}{*}{45}&S2,S10,S34,S52,S55,S63,S68,S86,S88,S89,\\
&S99,S100,S101,S123,S133,S140,S141,S162,S178\\
\hline
46-47&S21,S108,S160,S145\\
\hline
\end{tabular}}
\caption{Output for Protocol1 Based on Mean Values on 10 Trials (Random Sequence)}
\label{tb:200random}
\end{table}
\begin{table}[thbp]
{\footnotesize

\begin{tabular}{|c|l|}
\hline
\textbf{Mean Value}&\textbf{SUBTPAs}\\
\hline
36-37& S60, S158,S29,S82,S103,S147\\
\hline
38&S19,S48,S66,S108,S123,S157,S175\\
\hline
39&S52,S61,S84,S107,S113,S127,S141,S143,S149S167  \\
\hline
\multirow{2}{*}{40}& S33,S35,S37,S43,S72,S104,S106,\\
&S139,S159,S162,S166,S171,S185,S191\\
\hline
\multirow{3}{*}{41}& S3,S9,S15,S17,S20,S21,S23,S25,S26,S34,S39,S41,S42,\\
&S49,S76,S86,S90,S102,S117,S119,S121,S125,S130,S135,\\
&S156,S170,S172,S180,S186,S187,S190,S196,S197,S198\\
\hline
\multirow{3}{*}{42}&S5,S8,S11,S18,S22,S30,S36,S38,S44,S45,S55,S56,S57,\\
&S64,S77,S78,S79,S81,S83,S88,S100,S120,S128,S129,S132,\\
&S137,S138,S153,S154,S168,S169,S176,S182,S184,S194\\
\hline
\multirow{3}{*}{43}&S2,S4,S6,S7,S14,S16,S53,S54,S62,S63,S67,S69,S70,S71,\\
&S89,S91,S92,S98,S101,S105,S115,S118,S124,S134,S142,S144,\\
&S146,S150,S152,S160,S163,S165,S178,S179,S189,S195,S199\\
\hline
\multirow{2}{*}{44}&S1,S13,S24,S32,S51,S59,S68,S80,S85,S87,S94,S95,S96,\\
&S110,S112,S116,S131,S145,S151,S164,S183,S188,S193,S200\\
\hline
45&S27,S40,S47,S58,S73,S74,S114,S136,S140,S181\\
\hline
\multirow{2}{*}{46}&S10,S12,S28,S31,S46,S65,S93,S97,S109,\\
&S126,S133,S148,S155,S161,S173,S177,S192\\
\hline
47-49&S75,S111,S122,S50,S174,S99\\
\hline
\end{tabular}}
\caption{Output for Protocol1 Based on Mean Values on 10 Trials(Sobol Random Sequence)}
\label{tb:200sobol}
\end{table}

\newpage
Now, we compare the performance of protocol1 with Random sequence and Sobol Random Sequence and shown in Table~\ref{tb:200random} and Table~\ref{tb:200sobol} respectively. The tables has shown the average error detected by each SUBTPA in a ranked fashion and fit the least square lines for best fitting by using general equation shown in~(\ref{eq:ls1}) and~(\ref{eq:ls2}) respectively.

\begin{equation}\label{eq:ls1}
\left(\sum_{k=1}^N x_{k}^2\right)A + \left(\sum_{k=1}^N x_{k}\right)B= \sum_{k=1}^N x_{k} y_{k}
\end{equation}

\begin{equation}\label{eq:ls2}
\left(\sum_{k=1}^N x_{k}\right)A + NB= \sum_{k=1}^N y_{k}
\end{equation}

From the table~\ref{tb:200random} calculates the values and substitutes in the equations~(\ref{eq:ls1}) and~(\ref{eq:ls2}) and get the following linear equations in~(\ref{eq:lin1}) and~(\ref{eq:lin2}) as follows:
\begin{equation}\label{eq:lin1}
 2686700A+20100B=846216
\end{equation}
\begin{equation}\label{eq:lin2}
 20100A+200B=8415
\end{equation}

\begin{figure}[!htbp]
\centering
\includegraphics[height=3in,width=4.5in]{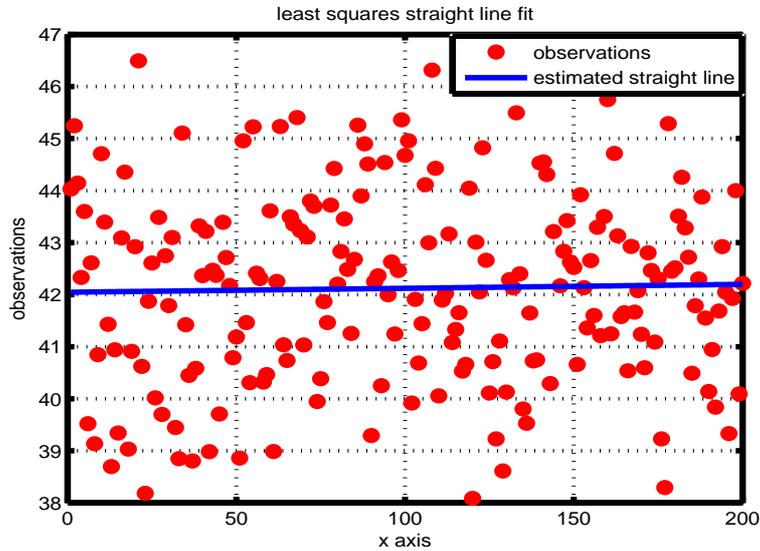}
\caption{Least Square line based on Mean error detection for 200 SUBTPA(Random Sequence) }
\label{fig:lstrandom}
\end{figure}

By solving the equations~(\ref{eq:lin1}) and ~(\ref{eq:lin2}) by using Cramer's rule we shall get the least square line for the best fit corresponding to the mean error detected by each SUBTPA when uses Random sequences in~(\ref{eq:leastlinerandom}) and the graph is in figure~\ref{fig:lstrandom}.
\begin{equation}
 y=Ax+B
\end{equation}
\begin{equation}\label{eq:leastlinerandom}
 y=0.00076x+41.99834
\end{equation}

\begin{figure}[!thbp]
\centering
\includegraphics[height=3in,width=4.5in]{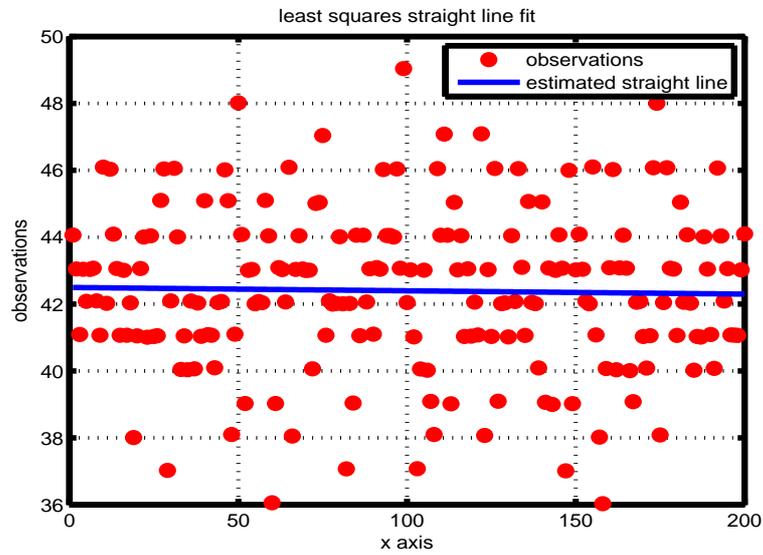}
\caption{Least Square line based on Mean error detection for 200 SUBTPA(Sobol Random Sequence)}
\label{fig:lstsobol}
\end{figure}

Similarly, from the table~\ref{tb:200sobol} calculates the values and substitutes in the equations~(\ref{eq:ls1}) and~(\ref{eq:ls2}) and get the following linear equations in~(\ref{eq:lin3}) and~(\ref{eq:lin4}) as follows:
\begin{equation}\label{eq:lin3}
 2686700A+20100B=850468
\end{equation}
\begin{equation}\label{eq:lin4}
 20100A+200B=8469
\end{equation}
By solving the equations~(\ref{eq:lin3}) and ~(\ref{eq:lin4}) by using Cramer's rule we shall get the least square line when uses Sobol Random sequences in equation~(\ref{eq:leastlinesobol}) and the graph is in figure~\ref{fig:lstsobol}.
\begin{equation}
 y=Ax+B
\end{equation}
\begin{equation}\label{eq:leastlinesobol}
 y=-0.001x+42.44548
\end{equation}

From, figure~\ref{fig:lstrandom} and figure~\ref{fig:lstsobol}, we can concludes that for sobol random sequence, points are much near to the best fit line but in case of random sequence the observed points are much more spread across over the plane. That indicates that for Sobol random sequence, error detected by each SUBTPA is approximately constant due to its uniform nature. Reversely, we can easily generates the mean error detected by each SUBTPA by using the equation~(\ref{eq:leastlinerandom})and~(\ref{eq:leastlinesobol}). Also, we can easily increases the SUBTPA size and change the parameters needs for construct equation in~(\ref{eq:ls1}) and~(\ref{eq:ls2}) and easily obtained the least square lines. By using the least square lines we can easily predict the average number of error detected by each SUBTPAs. Therefore, now by using least square line equation we can predict the mean error detected by each SUBTPA. Similarly, for protocol 2, protocol 3, and protocol 4 we can consruct the least square lines and measures the performance of our protocols. 
\chapter{Conclusions}
\label{ch:conc}
In this paper, we addressed the efficient Distributed Verification protocol based on the Sobol Random Sequence. We have shown that our protocol uniformly distributes the task among SUBTPAs. Most importantly, our protocol can handle failures of SUBTPAs due to its uniform nature and also gives better performance in case of unreliable communication link. Here, we mainly focussed on the uniform task distribution among SUBTPAs to detect the erroneous blocks as soon as possible. We used String Reconciliation Protocol to minimize the communication complexity between Coordinator and SUBTPA side. In addition, we reduce the workload at the Server side and also reduce the chance of network congestion at the Server side as well as Coordinator side by distributing the task. Thus, our Distributed Verification Protocol increases the efficiency and robustness of data integrity in Cloud Computing.


\appendix
\chapter{Root Finding of Polynomials}
\label{ch:app}
Assume we are given a polynomial $f(Z)$ of degree $m$ over a finite field $F_{q}$ .
This appendix briefly shows how to determine if all the zeros of $f(Z)$ are
distinct and lie in $F_{q}$ and, if so, how to find them using classical algorithms. The techniques described here are from~\cite{set} and are included here for completeness.

The particular type of root finding needed by the set reconciliation protocols involves three steps. First, determine if $f(Z)$ is square free. Second, verify that all irreducible factors of $f(Z)$ are linear. And finally, find the linear factors of $f(Z)$.

We can determine if $f(Z)$ is square-free by computing the GCD (great-
est common divisor) of $f(Z)$ and its derivative $f'(Z)$ using the Euclidean
algorithm in $O(m^{2} )$ field operations. To verify that $f(Z)$ is the produce
of $m$ linear factors, we simply verify that $f(Z) = GCD(f(Z), Z ^{q}-Z)$, the
latter term being the product of all monic linear polynomials over $F_{q}$ . This
verification can be completed in $O(m^{2}\; log\; q)$ time by using repeated squaring
$(mod\; f (Z))$, giving an overall verification time of $O(m^{2}\; log\; q)$.

Finally, we find the linear factors of $f(Z)$ using probabilistic techniques.
We consider two different cases for the field $F_{q}$ (corresponding to the possible
choices for use in our set reconciliation protocols): one where $q$ is a prime
and the other where $q = 2^{b}$ . When $q$ is a prime, note that the elements of
$F_{q}$ are zeros of 

\begin{displaymath}
 Z^{q}-Z=(Z^{\frac{q-1}{2}}+1) \cdot Z\cdot (Z^{\frac{q-1}{2}}-1)
\end{displaymath}
So, almost half of the elements of $F_{q}$ are zeros of $R(Z) = Z^{\frac{q-1}{2}}-1$.

A polynomial with similar properties can also be constructed for the field $F_{2^{b}}$:
$$
R(Z)= Z^{2^{b-1}}+Z^{2^{b-2}}+\cdots+Z^{4}+Z^{2}+Z
$$
We then have that
\begin{equation*}
\begin{split}
R(Z)\cdot (R(Z)+1) &= R(Z)^{2}+R(Z)\\
                  &=Z^{2^{b}}+Z^{2^{b-1}}+\cdots +Z^{2}+R(Z)\\
                  &=Z^{2^{b}}+Z
\end{split}
\end{equation*}
So, all the elements of $F_{2^{b}}$ are zeros of $R(Z)\cdot(R(Z)+1)$, and each element
is either a zero of $R(Z)$ or of $R(Z) + 1$.

To determine the zeros of $f(Z)$, we chose a random element of $a\in F_{q}$ and compute $GCD(f(Z),R(Z - a))$, which will have almost half the degree of $f(Z)$. Applying this technique recursively on the two factors of $f(Z)$, with different values for a will further split the polynomial, ultimately into linear factors. In total, the expected number of GCDs required will be $O(d)$.


\end{document}